\documentclass[12pt,preprint]{emulateapj}
\usepackage{apjfonts}

\submitted{Accepted for Publication in The Astrophysical Journal}

\shorttitle{Full Polarization Spectra}
\shortauthors{Homan et al.}

\begin{document}

\title{Full Polarization Spectra of 3C\,279}
\author{D. C. Homan\altaffilmark{1}, M. L. Lister\altaffilmark{2}, 
H. D. Aller\altaffilmark{3}, M. F. Aller\altaffilmark{3}, 
J. F. C. Wardle\altaffilmark{4}} 

\altaffiltext{1}{Department of Physics and Astronomy, Denison University, 
Granville, OH  43023; homand@denison.edu}

\altaffiltext{2}{Department of Physics, Purdue University, 525 Northwestern
Avenue, West Lafayette, IN 47907; mlister@physics.purdue.edu}

\altaffiltext{3}{Department of Astronomy, University of Michigan, Ann Arbor,
MI 48109; haller@umich.edu, mfa@umich.edu}

\altaffiltext{4}{Department of Physics, Brandeis University,
Waltham, MA 02454; wardle@brandeis.edu}

\begin{abstract}
We report the results of parsec-scale, multi-frequency VLBA
observations of the core region of 3C\,279 in Stokes $I$,
linear polarization, and circular polarization.  These
full polarization spectra are modeled by radiative
transfer simulations to constrain the magnetic field and 
particle properties of the parsec-scale jet in 3C\,279.
We find that the polarization properties of the core
region, including the amount of linear polarization, the
amount and sign of Faraday rotation, and
the amount and sign of circular polarization
can be explained by a consistent physical picture.  The
base of the jet, component D, is modeled as an
inhomogeneous Blandford-K\"onigl style conical jet dominated by a 
vector-ordered poloidal magnetic field along the jet axis, and we  
estimate its net magnetic flux.  This poloidal field 
is responsible for the linear and circular polarization from 
this inhomogeneous component.
Farther down the jet the magnetic field in two homogeneous 
features is dominated by local shocks and a smaller fraction
of vector-ordered poloidal field remains along the jet axis.  
This remaining poloidal field provides  
internal Faraday rotation which drives Faraday conversion
of linear polarization into circular polarization from 
these components.  
In this picture, we
find the jet to be kinetically dominated by protons with 
the radiating particles being dominated by electrons at an
approximate fraction of $\gtrsim 75$\%, still allowing the 
potential for a significant admixture of positrons. 
Based on the amounts of Faraday conversion deduced
for the homogeneous components, we find a plausible range 
for the lower cutoff in the relativistic particle energy spectrum to be 
$5 \lesssim \gamma_l \lesssim 35$.
The physical picture described here is not unique if the 
observed Faraday rotation and depolarization occur in screens 
external to the jet; however, we find the joint explanation 
of linear and circular polarization observations from a 
single set of magnetic fields and particle properties internal 
to the jet to be compelling evidence for this picture.    
\end{abstract}


\section{Introduction}
\label{s:intro}
The three-dimensional magnetic field structures and the particle populations 
of extragalactic jets from Active Galactic Nuclei (AGN) are still not well understood.  
\citet{LB02} and \citet{LCB06} have recently made progress in studying the
3-D magnetic field structures of kilo-parsec scale jets, but little is known
about the 3-D field structure near the jet origin, on parsec or sub-parsec
scales.  We wish to know if the jet magnetic field shows a structure which
has its roots in the magnetic field in the supermassive black-hole/accretion-disk 
system responsible for giving rise to the jets \citep[e.g.][]{BZ77,M01,K02,VK04}. 
\citet{M02} has
suggested a direct disk-jet connection on the basis of X-ray/radio correlations,
and this connection may extend to the magnetic field which threads the accretion
disk around the black hole.  Unanswered questions include, for example, is there 
significant vector-ordered
poloidal field along the jet axis or perhaps a toroidal/helical field structure confining
the jet and indicating a jet current \citep{A02,G04,F06}? 
It is also unknown whether the particle population
of jets is primarily electron-proton, $p^+e^-$, or electron-positron, $e^+e^-$ 
\citep[e.g.][]{WHOR98,SM00,D06}.  We parameterize this unknown as the  
lepton number, $\ell = (n_- - n_+)/(n_-+n_+)$, where $n_-$ and $n_+$ are the
number densities of electrons and positrons respectively.  Additionally, the
limits on the power-law particle spectrum which gives rise to the observed
synchrotron radiation are poorly constrained.  The relativistic number density
can be parameterized as $N_\gamma d\gamma = K\gamma^{-p} d\gamma$ 
for $\gamma_l \leq \gamma \leq \gamma_u$ where it is assumed to have a
hard cutoff to the power-law at both high and low energies. \citet{CF93} found that the
lower cutoff $\gamma_l$ set the scale for the bulk kinetic 
luminosity of jets because the low energy particles dominate the particle
density. 

With sub-milliarcsecond resolution, Very Long Baseline Array (VLBA) 
observations of extragalactic radio jets can study synchrotron emission 
from jets within a few parsecs of the jet origin.  Measurements of linear polarization from
milli-arcsecond jets are sensitive to the net magnetic field order in the plane
of the sky; however, multiple 3-D magnetic field structures can all yield the same
observed linear polarization.  For example, a transverse shock of tangled magnetic field, 
a toroidal field, or a high-pitched helical field will all produce linear 
polarization with the electric field vector parallel to the jet axis.  Likewise, 
a vector-ordered poloidal field, shear in a tangled field, or a low-pitch helical 
field will all produce linear polarization with the electric field vector transverse 
to the jet axis.  Faraday rotation of linear polarization
is sensitive to magnetic field order along the line of sight and to the properties
of the particles doing the rotation. The observed Faraday rotation must, in some
cases, result from thermal particles and magnetic fields which are along
the line of sight but external to the jet, but for rotations smaller than about 
$45^\circ$, it is difficult to distinguish internal from external Faraday 
rotation without additional information \citep{B66}.  Recently, several authors
have attributed Faraday rotation gradients observed approximately transverse to
jets as evidence of toroidal or helical magnetic fields either in the jet itself
or in a sheath layer around the jet \citep{A02,G04,ZT05,A05,GMJA08}; however, these
gradients are often gradients in magnitude of a single sign of the observed rotation 
and do not show the clear anti-symmetric signature expected for such fields, so an 
additional external Faraday screen must be proposed as well.  Alternatively
the observed gradients in rotation measure could be due to density gradients 
in the material surrounding the jet.  
In the case of 3C\,273, the case for toroidal or helical fields 
is stronger as the same direction of the gradient is observed at multiple 
jet locations \citep{ZT05,A05,A08}.  \citet{MJ08} have also argued for
helical field structure in BL Lacertae based on temporal rotations 
in linear polarization angle from the jet core region. 

Parsec-scale observations of circular polarization from extra-galactic jets 
provide additional constraints that can break some of the degeneracies 
inherent in linear polarization observations.  Circular polarization may 
be generated either as an {\em intrinsic} component of the emitted synchrotron
radiation or via Faraday {\em conversion} of linear polarization into circular
polarization \citep[e.g.][]{Jones88}.  Like Faraday rotation, {\em conversion} is a 
bi-refringence effect; however, unlike rotation, conversion is much 
stronger in relativistic particles than thermal particles, and we don't
expect significant Faraday conversion from magnetic fields and particles external
to the jet \citep{JOD77,HAW01}.  In this way, circular polarization probes
the jet magnetic fields and particles directly without modification from 
external screens.

Parsec-scale circular polarization observations of AGN jets have been reported by 
\citet{WHOR98,HW99,HAW01,HW03,HW04,HL06}, and most recently by \citet{G08}. 
Most AGN jets appear to have less than $\sim 0.1$ to $0.2$\% circular 
polarization, with 10\% to 20\% of jets detected at the level of $\sim 0.3$\% 
to $1.0$\% of the Stokes $I$ emission from or very near the base of the jet (or ``jet core'').  
The highest levels of circular polarization detected are $2$ to $4$\% 
of the local Stokes $I$ emission in the core region of the nearby 
radio galaxy 3C\,84 \citep{HW04} and in the intra-day variable 
source PKS 1519$-$273 \citep{MKRJ00}.  Single dish monitoring of circular 
polarization from AGN jets by the UMRAO has been ongoing since 2002 \citep{AAP03}, 
and the Austrian Compact Telescope Array has studied the integrated circular 
polarization from AGN \citep{RNS00}.  Circular polarization has also been 
observed in intra-day variable sources, micro-quasars, low luminosity
AGN, and the Galactic Center \citep{BFB99,BFM02,BBFM01,F00,F02,MKRJ00,SM99}.  
To date, no strong 
correlations have been found between the appearance of circular 
polarization and other source properties \citep{HAW01, RNS00, HL06}; 
however, the lack of strong correlations may simply be related to the small
fraction of detected sources and the
low levels of circular polarization in those objects.  There is increasing evidence 
that at least some circularly polarized sources tend to have a preferred ``handedness'' 
of circular polarization \citep{K84,HW99,HAW01,BFM02,HL06} suggesting
a persistent magnetic field structure is responsible for setting the sign
of the observed polarization, although it is important to note that 
changes in sign have been observed in some cases \citep[e.g.][]{AAP03}.

To date, radiative transfer modeling of parsec-scale circular 
polarization has been at only one or two frequencies, limiting
our ability to uniquely constrain physical jet properties.  
\citet{WHOR98} preferred a model for 3C\,279 where circular
polarization was produced via Faraday conversion of linear into 
circular polarization and this conversion was driven by some
internal Faraday rotation.  Their results implied a low cutoff in the 
relativistic particle spectrum of $\gamma_l \leq 20$, suggesting
the jet was dominated by electron-positron pairs to keep the jet's
kinetic luminosity within reasonable bounds \citep[e.g.,][]{CF93}.
However, subsequent work by \citet{RB02} and \citet{BF02} showed
that some addition of thermal matter to the jet allowed a larger
value for $\gamma_l$, reducing the need for a electron-positron 
dominated jet.  \citet{HW03} and \citet{HW04} also preferred
Faraday conversion models for PKS 0607$-$157 and 3C\,84 respectively.  
\citet{G08} have linked the observation of transverse rotation 
measure gradients with parsec-scale circular polarization observations
in a qualitative fashion to show that both observations are 
consistent with helical magnetic fields if all eight jets
in their sample emerge from the south magnetic pole of the
central engine.  

To make progress on using parsec-scale circular and linear 
polarization observations of AGN jets to constrain their magnetic 
field structure and particle populations, we have embarked on
a program to study the linear polarization, circular polarization,
and Stokes $I$ spectra of AGN jets at six frequencies with the VLBA.  
We call these ``full polarization spectra'' of AGN jets, as we use all
four Stokes parameters to constrain radiative transfer models. 
We combine these single-epoch, parsec-scale data with integrated monitoring
by the UMRAO at 4.8, 8.0, and 14.5 GHz.  The first VLBA observations 
in this program took place in November of 2005 for 18 jets.  We detected 
strong circular polarization at multiple frequencies in three jets: 3C\,84, 
3C\,279, and 3C\,380. Here we report our results on 3C\,279.  Future papers 
will explore 3C\,84 and 3C\,380, as well as the sources in our sample
with little or no detected circular polarization.   In \S{\ref{s:obs}}
we describe our observations, calibration, and Gaussian modeling 
of the core region of 3C\,279 to obtain spectra for fitting.  In \S{\ref{s:rad}} 
we describe a variety of possible models for producing circular and linear 
polarization in jets and then use analytic and computational radiative 
transfer models to fit the full polarization spectra obtained in \S{\ref{s:obs}}.
We discuss the results of the radiative transfer modeling in \S{\ref{s:discuss}}, 
and our conclusions appear in \S{\ref{s:conclude}}.  Throughout the paper
we assume a cosmology where $H_o = 70$ km/s/Mpc, $\Omega_M = 0.3$, and 
$\Omega_\Lambda = 0.7$, and so at the redshift of 3C279 ($z = 0.538$), 
one milli-arcsecond corresponds to a projected linear scale of 6.34 parsecs.

\section{Observations}
\label{s:obs}

\subsection{VLBA Observations and Reduction}

In 2005, November 17$-$20, we observed 18 AGN radio jets with the 
National Radio Astronomy Observatory's Very Long Baseline Array 
(VLBA)\footnote{The National Radio Astronomy Observatory is a facility 
of the National Science Foundation operated under cooperative 
agreement by Associated Universities, Inc.}
at six frequencies: 8.01 GHz, 8.81 GHz, 12.35 GHz, 15.37 GHz, 22.23 GHz, 
and 24.35 GHz.  The observations were recorded at each frequency in 
dual circular polarization with single-bit recording at four intermediate 
frequencies (IFs) of 8 MHz bandwidth for a total bandwidth of 32 MHz.  
The observations were made over a continuous 72 hour period to minimize 
the effects of source variability with each of three 24 hour segments 
devoted to a pair of frequencies: 8.0 and 22 GHz, 12 and 15 GHz, and 8.8 and 
24 GHz respectively.  Each source was visited 9 times at each observing 
frequency with these ``scans'' highly interleaved with neighboring 
sources to maximize (u,v)-plane and parallactic angle coverage.  
Scan lengths were approximately 300 seconds at 22 and 24 GHz, 215 seconds 
at 12 and 15 GHz, and 130 seconds at 8.0 and 8.8 GHz.  Each source thus obtained 
approximately 45 minutes integration time per frequency at 22 and 24 GHz,
32 minutes at 12 and 15 GHz, and 19 minutes at 8.0 and 8.8 GHz. Our calibrated
Stokes I and polarization images at each of these frequency bands appear
in Figures 1, 2, and 3 respectively.   

\begin{figure*}
\figurenum{1}
\begin{center}
\includegraphics[scale=0.9,angle=0]{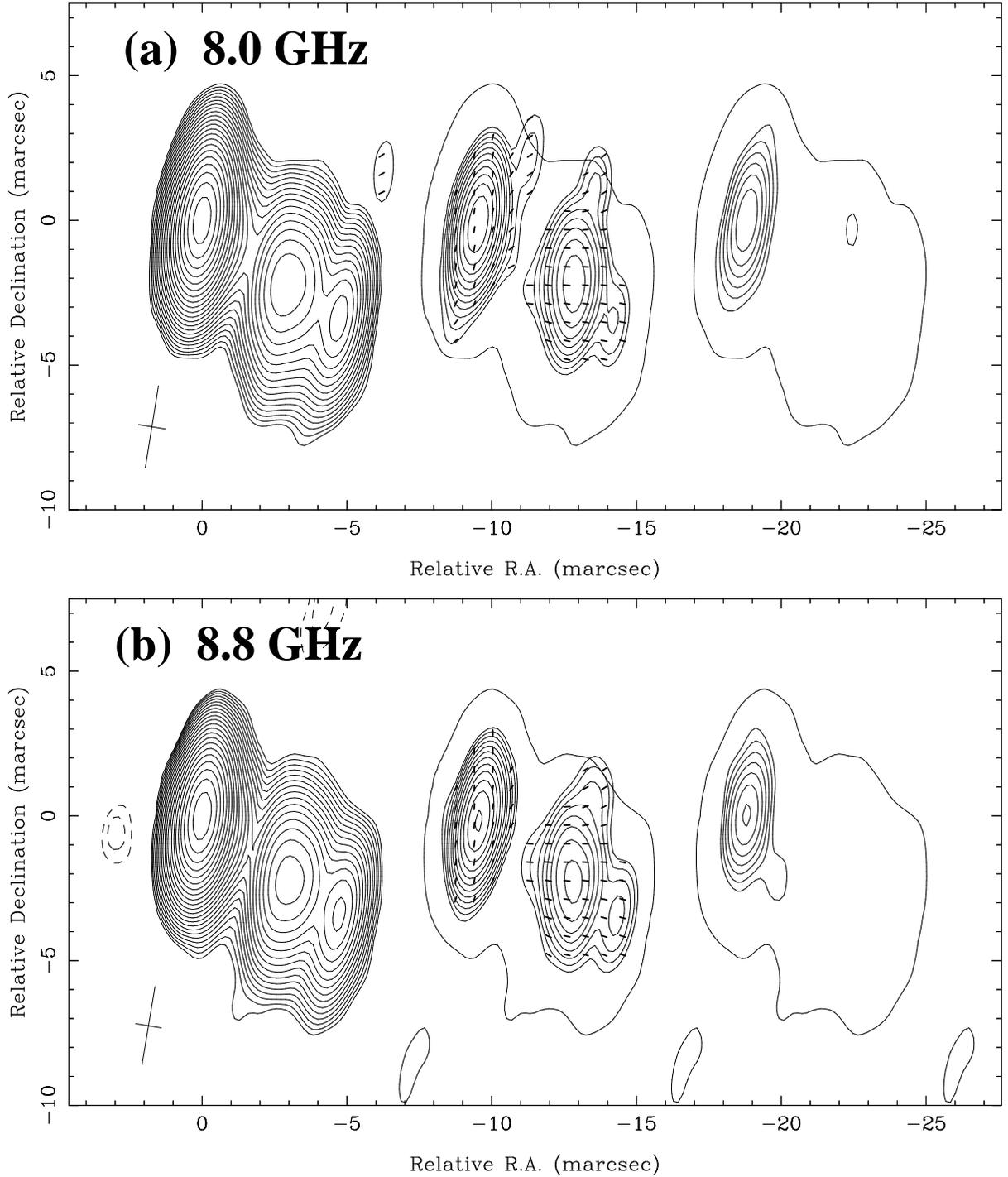}
\end{center}
\figcaption[]{
Contour images of Stokes-I flux density (left), linearly polarized flux density (middle) 
and circularly polarized flux density (right) at 8.0 GHz (panel (a)) and 8.8 GHz (panel (b)).  
Contour levels begin at 5 mJy/beam for Stokes-I and circular polarization, and 10 mJy/beam 
for linear polarization, and increase in steps of $\times\sqrt{2}$.  Tick marks on the 
linear polarization image represent the measured electric vector position angle.  A single
contour from the Stokes-I image bounds the linear and circular polarization images to
show registration. A cross-figure representing the FWHM dimensions of the restoring beam
appears in the lower left hand corner of each panel.
}
\end{figure*}

\begin{figure*}
\figurenum{2}
\begin{center}
\includegraphics[scale=0.85,angle=0]{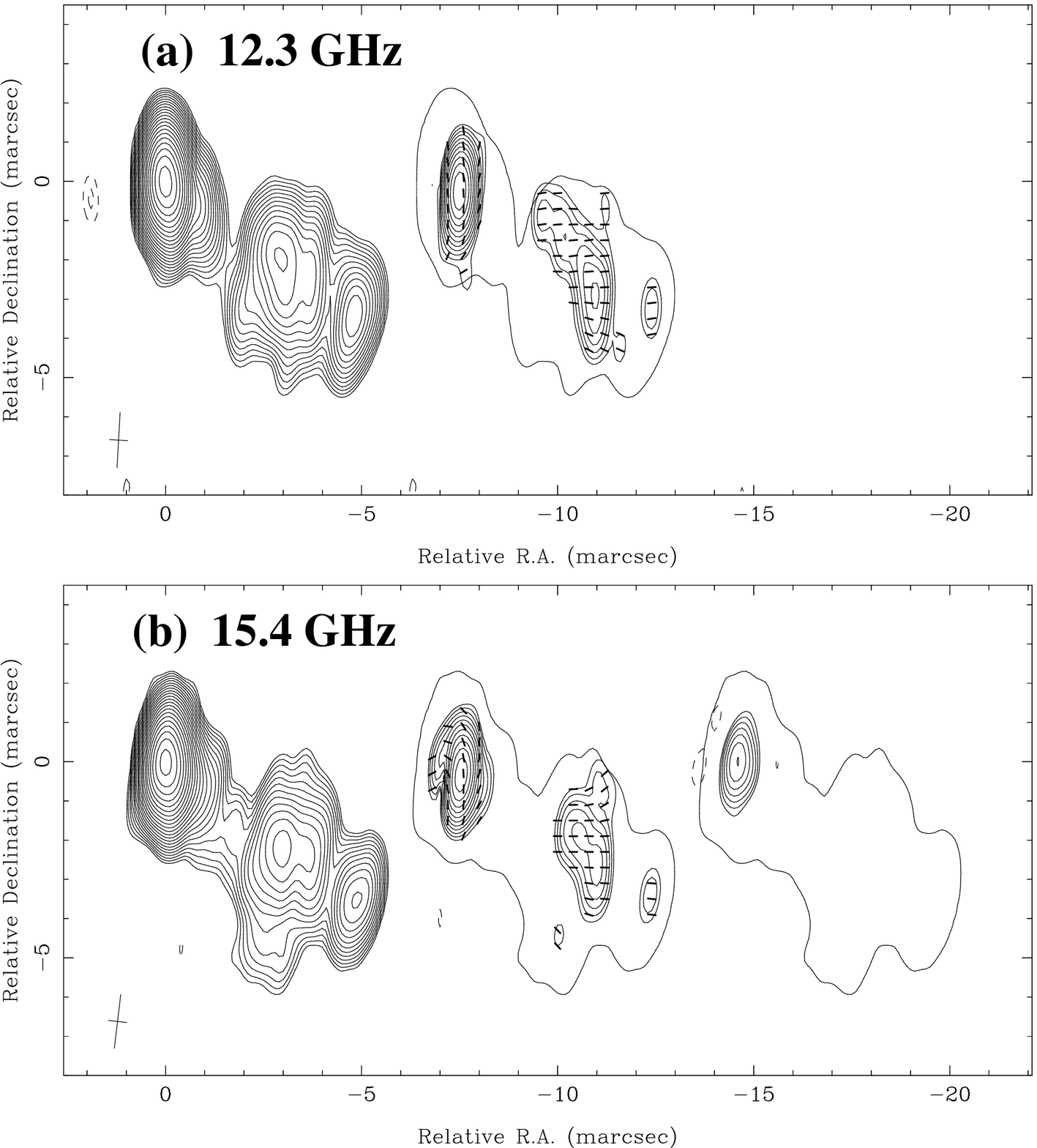}
\end{center}
\figcaption[]{
Contour images of Stokes-I flux density (left), linearly polarized flux density (middle) 
and circularly polarized flux density (right) at 12.3 GHz (panel (a)) and 15.4 GHz (panel (b)).  
Contour levels begin at 5 mJy/beam for Stokes-I and circular polarization, and 10 mJy/beam 
for linear polarization, and increase in steps of $\times\sqrt{2}$.  Tick marks on the 
linear polarization image represent the measured electric vector position angle.  A single
contour from the Stokes-I image bounds the linear and circular polarization images to
show registration. Note that severe RFI at 12.3 GHz prevented reliable measurement of 
the circular polarization at this frequency, so those data are not presented, see 
\S{\ref{s:obs}} for a more detailed description of this problem.
A cross-figure representing the FWHM dimensions of the restoring beam
appears in the lower left hand corner of each panel.
}
\end{figure*}

\begin{figure*}
\figurenum{3}
\begin{center}
\includegraphics[scale=0.9,angle=0]{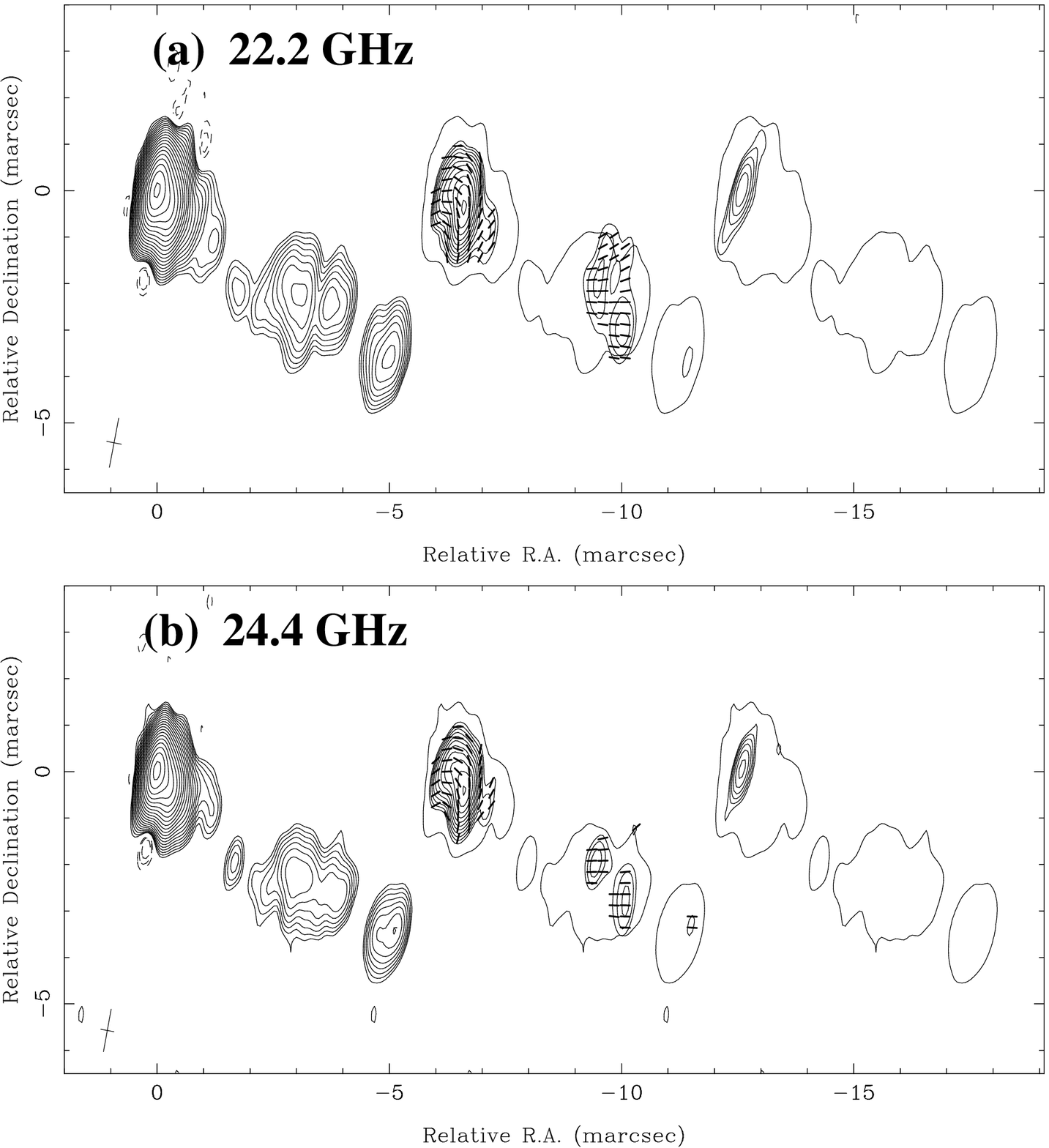}
\end{center}
\figcaption[]{
Contour images of Stokes-I flux density (left), linearly polarized flux density (middle) 
and circularly polarized flux density (right) at 22.2 GHz (panel (a)) and 24.4 GHz (panel (b)).  
Contour levels begin at 10 mJy/beam and increase in steps of $\times\sqrt{2}$.  Tick marks on the 
linear polarization image represent the measured electric vector position angle.  A single
contour from the Stokes-I image bounds the linear and circular polarization images to
show registration. 
A cross-figure representing the FWHM dimensions of the restoring beam
appears in the lower left hand corner of each panel.
}
\end{figure*}

A-priori data calibration, fringe-fitting, self-calibration, and calibration
for linear and circular polarization were performed at Denison University
using the techniques described in detail by \citet{LH05} and \citet{HL06}. 
At 15.4 GHz, the electric vector position angle (EVPA) of the linear 
polarization was calibrated by comparing to the MOJAVE database of rotated 
D-terms, as described in \citet{LH05}.  At the other frequencies, we used the
polarization of eight well defined jet components and looked at their deviations
from the 15 GHz calibration (which was taken as absolute); although there was
some clear Faraday Rotation in three of the components, the offsets at each
frequency were remarkably consistent.  We did make corrections for the
apparent rotation measures of those three components, but our calibration is
not strongly dependent on those values.  Indeed, if we rely on only the five
components without apparent Faraday rotation, we get the same calibration to
within a degree at each frequency.  Based on the scatter of the offsets 
between components, we estimate that our EVPA calibration is good to better
than one degree at 12 GHz and below, and good to better than two degrees at 22
and 24 GHz.  These numbers are relative to the 15 GHz calibration, which we 
expect to be accurate to within one degree, \citep{LH05}; however, relative 
calibration between the frequencies is our main concern here for the modeling 
of the emission region.  In our results we therefore consider only calibration uncertainty
relative to 15 GHz added in quadrature with modeling uncertainty.

Circular polarization calibration was done via the gain transfer technique described
in detail by \citet{HL06}, including their Monte Carlo methods to estimate the
final uncertainty in the observed circular polarization.  We have made some
small improvements upon the methods described there to allow 
semi-automatic flagging of antennas and some individual scans that show a 
high RCP/LCP gain ratio.  In particular we flagged an antenna if the standard 
deviation of its RCP/LCP antenna gain ratio was more than twice the median standard 
deviation of all the antennas.  This requirement only resulted in the flagging 
of the OV antenna at 8.0 GHz which had a very large (factor of $\simeq 2$) gain discrepancy 
between the right and left hand feeds.  We also flagged individual scans 
if their RCP/LCP ratio deviated by more than 5 times the standard deviation
from the smoothed gains.  In practice this accounted for a very small 
percentage, $0.1-0.3$ percent, of the overall scans being flagged.  As 
described in \citet{HL06}, we also excluded sources with low SNR, as
indicated by large RCP/LCP gain fluctuations, or with apparently high levels of 
circular polarization, as indicated by a larger RCP/LCP gain offset from the
smoothed gains, from contributing to the final smoothed gain-transfer table which 
would determine the overall circular polarization calibration.\footnote{The 
formal limit on the RCP/LCP gain fluctuations for low SNR sources was if 
the source had a standard deviation larger than twice the median standard 
deviation of all sources.  The formal limit for a source with large apparent
circular polarization was 0.5\%, as indicated by a systematic RCP/LCP gain offset
of $\geq 0.005$ from the smoothed gains.}  These requirements excluded
only between one and three sources from contributing to the 
final gain smoothing at each frequency, except at 24 GHz where seven sources 
were excluded, likely due to lower overall SNR at that frequency. Our circular 
polarization observations at 12 GHz were corrupted by strong radio frequency interference
(RFI), likely from geostationary satellites passing within the telescope beam, 
so we do not report circular polarization results at that frequency.
 
As discussed in detail in \citet{HL06}, phase calibration for circular 
polarization is complicated if a source has significant extended structure
in Stokes $I$.  This may lead to spurious anti-symmetric 
structure in the resulting circularly polarized images due to phase errors
in earlier steps in the calibration.  To account for
this effect, we use the same approach described and tested by \citet{HL06} 
of adding an extra round of phase self-calibration, assuming no circular 
polarization in the data.  In general this extra round of phase 
self-calibration is very effective in removing genuine phase 
gain errors while preserving the original circularly polarized signal; 
although \citet{HL06} did find that the amplitude of the circularly polarized
signal may be reduced by a few percent up to 10\% and the position of
the circular polarization was shifted a small amount, less than about half
a beam-width, to better align with the source peak. Real
circularly polarized structure significantly away from the source peak
was not shifted by the procedure.  In the case of a source with a broad 
core region, \citet{HL06} found that the circular polarization may be spread by 
this procedure to encompass the whole region.    

In recognition of the potential uncertainty of the precise placement of
the circular polarization within the core region of 3C279, we measured the 
integrated amount of circular polarization in the core region at each frequency 
using a single Gaussian component in the (u,v)-plane, and these results are 
reported in Table 1.  

\begin{deluxetable}{rc|c}
\tablecolumns{3}
\tablewidth{0pc}
\tabletypesize{\scriptsize}
\tablenum{1}
\tablecaption{Circular Polarization of Core Region.\label{t:core_cp}}
\tablehead{\multicolumn{2}{c}{\underline{Measured Values}} & 
\colhead{\underline{If All Stokes-V on ``D''}} \\
\colhead{$Freq.$} & \colhead{$V$} & \colhead{$m_c$} \\ 
\colhead{(GHz)} & \colhead{(mJy)} & \colhead{(\%)} \\
\colhead{(1)} & \colhead{(2)} & \colhead{(3)}}
\startdata
$  8.01$ & $ 50.8 \pm  9.6$ & $  1.99 \pm  0.38$ \\ 
$  8.81$ & $ 48.2 \pm  8.0$ & $  1.55 \pm  0.26$ \\ 
$ 12.35$ & \nodata & \nodata \\ 
$ 15.37$ & $ 47.6 \pm  7.9$ & $  0.88 \pm  0.15$ \\ 
$ 22.23$ & $ 58.2 \pm 13.4$ & $  0.86 \pm  0.20$ \\ 
$ 24.35$ & $ 71.0 \pm 16.5$ & $  0.94 \pm  0.22$ \\ 
\enddata
\tablecomments{
Columns are as follows: 
(1) Frequency of observation in GHz; 
(2) Stokes-V flux density for the core region in mJy;
(3) Fractional circular polarization under the hypothetical scenario, 
discussed in \S{3.4.1}, where all the Stokes-V flux is associated with 
component ``D'' in the core region.
}
\end{deluxetable}

As described below, the core region of 3C279 is well modeled in Stokes $I$ and
linear polarization by three closely spaced components labeled D, 5, and 4 
in Figure 4, and the third column of Table 1 shows the fractional
circular polarization at each frequency if all of the measured circular
polarization originated from component D.  However, as described above, the 
complications of phase calibration for circular polarization limit our 
ability to confidently divide the measured circular 
polarization between these components at the lower frequencies, and indeed
at 8.0 and 8.8 GHz, the measured circular polarization could come from
any combination of these three components.  At 15 GHz, the circular 
polarization is most likely associated with some combination of D and 5.
At 22 and 24 GHz, we are confidently able to assign the measured circular
polarization to component D at the base of the jet.  To confirm this, we
ran tests of our phase calibration procedure at 22 and 24 GHz by
generating simulated data with the same Stokes $I$ structure as the clean component
models at these frequencies, but with the addition of a 10 or 20 mJy circularly 
polarized component at the locations of components 5 or 4 with the remaining
circular polarization on component D.  We found that the extra round of 
phase-calibration assuming zero circular polarization was not able to move 
circular polarization that originated on components 5 or 4 to component D.
These results give us confidence that the circular polarization observed on
component D indeed originated on component D and was not transferred there
by our phase calibration.  

\begin{figure}
\figurenum{4}
\includegraphics[scale=0.45,angle=0]{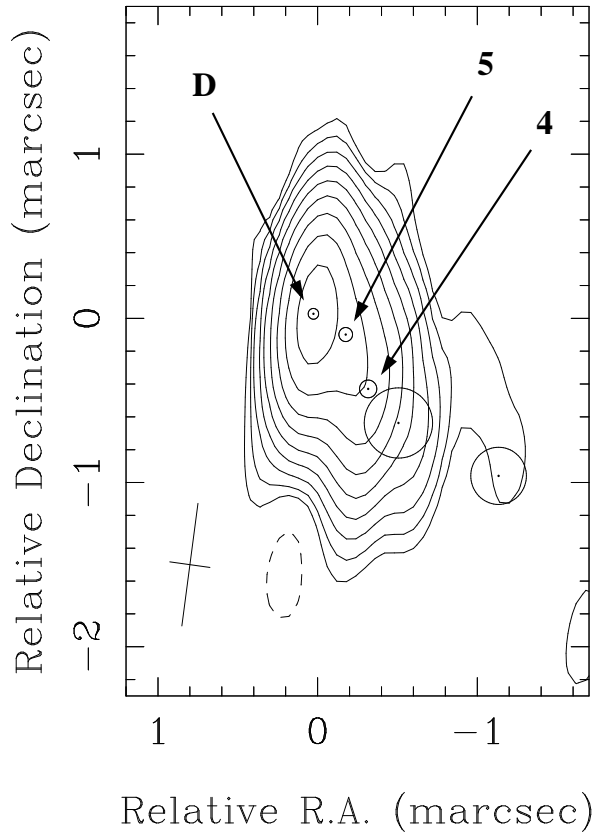}
\figcaption[]{
Uniformly weighted contour image of the core region at 24 GHz with the main core 
components labeled. Note that component positions and FWHM sizes plotted here are 
from a single possible
set of model components.  As described in the text, we fit seven possible sets 
of components to span a range of plausible scenarios for fitting the structure of
this source.  For example, the two weaker, unlabeled components in this figure 
are fit as a single component in two of our seven possible models.
}
\end{figure}

\subsection{Gaussian Model of the Core Region}

We used the Caltech VLBI program {\em Difmap} \citep{S97} to model the
Stokes $I$ structure and linear polarization of 3C279 at each frequency
with particular attention to the structure of the core region which
produces the observed circular polarization. One challenge to this process 
is that the angular resolution of our observations differs by a factor of three between
8.0 and 24 GHz.  To fit a comparable model over this range of resolution, we
required that the relative separation, sizes, and shapes of the Gaussian 
components remain the same from one frequency to another; however, we allowed
the fluxes of each component to vary as well as the overall phase center of
the model.  

One difficulty with this approach is that no single collection of 
model components will be ideal for all frequencies, so in total we repeated
this process to fit seven separate sets of model components in this fashion.
Each of these separate sets of model components explored somewhat different 
assumptions for how the structure of the core region and the remainder of the 
source was fit. For example, in some cases we fit elliptical Gaussians to the core
region and in other cases circular Gaussian.  We also tried varying the 
relative positions of the core components as well as the properties
and numbers of components used to describe the rest of the jet.   
Taken together, these seven different sets of model components sampled a range 
of reasonable possibilities for fitting the structure of 3C279.  
Overall the results for the core region were quite similar from one possibility 
to the next, and in Table 2 we report the average fluxes and polarizations from 
these different possibilities for each of the main core components.  The uncertainties for
the component properties given in Table 2 include calibration 
uncertainties\footnote{The overall flux calibration at each frequency was assumed 
to be good to 5\% based on the results reported in the appendix of 
\citet{H02}.}
added in quadrature with the standard deviation of the component values from
the seven different sets of model components.  Figure 4 shows the size
and spacing of the three core region components from a representative 
set of model components.

\begin{deluxetable}{crrrr}
\tablecolumns{5}
\tablewidth{0pc}
\tabletypesize{\scriptsize}
\tablenum{2}
\tablecaption{Flux and Linear Polarization of Core Region Components.\label{t:comp_prop}}
\tablehead{\colhead{Component} & \colhead{$Freq.$} & \colhead{$I$} & 
\colhead{$m_l$} & \colhead{$\chi$}\\
\colhead{ID} & \colhead{(GHz)} & \colhead{(Jy)} & \colhead{(\%)} & \colhead{(degrees)}\\
\colhead{(1)} & \colhead{(2)} & \colhead{(3)} & \colhead{(4)} &
\colhead{(5)}}
\startdata
  D & $  8.01$ & $ 2.553 \pm  0.301$ & $  3.3 \pm  0.7$ & $ 118.9 \pm  4.0$ \\ 
    & $  8.81$ & $ 3.118 \pm  0.304$ & $  3.9 \pm  0.5$ & $ 113.1 \pm  1.7$ \\ 
    & $ 12.35$ & $ 4.660 \pm  0.441$ & $  3.0 \pm  0.3$ & $ 118.7 \pm  2.7$ \\ 
    & $ 15.37$ & $ 5.406 \pm  0.393$ & $  2.5 \pm  0.2$ & $ 112.6 \pm  3.1$ \\ 
    & $ 22.23$ & $ 6.759 \pm  0.431$ & $  2.4 \pm  0.1$ & $ 111.5 \pm  3.1$ \\ 
    & $ 24.35$ & $ 7.531 \pm  0.476$ & $  2.2 \pm  0.1$ & $ 100.9 \pm  2.8$ \\ 
  5 & $  8.01$ & $ 3.914 \pm  0.355$ & $  5.4 \pm  0.8$ & $  18.1 \pm  5.4$ \\ 
    & $  8.81$ & $ 3.865 \pm  0.354$ & $  7.5 \pm  0.9$ & $  19.9 \pm  1.9$ \\ 
    & $ 12.35$ & $ 3.944 \pm  0.331$ & $  6.8 \pm  0.8$ & $  37.4 \pm  2.4$ \\ 
    & $ 15.37$ & $ 3.329 \pm  0.246$ & $  8.6 \pm  0.6$ & $  41.6 \pm  1.8$ \\ 
    & $ 22.23$ & $ 2.812 \pm  0.236$ & $ 10.0 \pm  0.9$ & $  49.3 \pm  2.2$ \\ 
    & $ 24.35$ & $ 2.977 \pm  0.260$ & $  9.0 \pm  0.8$ & $  48.8 \pm  2.0$ \\ 
  4 & $  8.01$ & $ 2.964 \pm  0.433$ & $ 13.3 \pm  1.7$ & $ 163.1 \pm  3.3$ \\ 
    & $  8.81$ & $ 2.817 \pm  0.397$ & $ 13.5 \pm  1.3$ & $ 164.4 \pm  2.1$ \\ 
    & $ 12.35$ & $ 3.178 \pm  0.313$ & $ 13.5 \pm  0.6$ & $ 174.3 \pm  1.5$ \\ 
    & $ 15.37$ & $ 2.668 \pm  0.228$ & $ 16.3 \pm  1.0$ & $ 175.6 \pm  1.2$ \\ 
    & $ 22.23$ & $ 2.235 \pm  0.157$ & $ 16.6 \pm  0.8$ & $ 181.2 \pm  2.2$ \\ 
    & $ 24.35$ & $ 2.211 \pm  0.160$ & $ 16.3 \pm  0.8$ & $ 179.3 \pm  1.9$ \\ 
\enddata
\tablecomments{
Columns are as follows: (1) Component identifier; 
(2) Frequency of observation in GHz; 
(3) Stokes-I flux density in Jy;
(4) Percent fractional linear polarization;
(5) Electric vector position angle of linear polarization. 
}
\end{deluxetable}

\subsection{UMRAO Integrated Monitoring of 3C279}

\begin{figure}
\figurenum{5}
\hspace{-0.4in}
\includegraphics[scale=0.37,angle=-90]{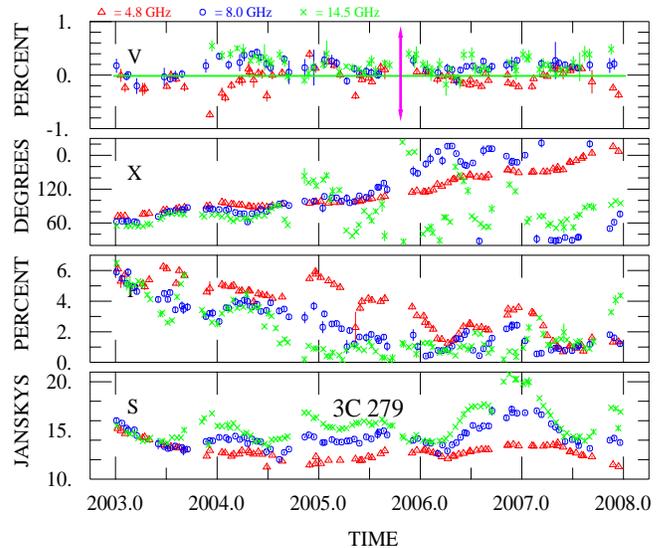}
\figcaption[]{
\label{f:UMRAO}
Intergrated polarization monitoring by the UMRAO at 4.8, 8.0, and 14.5 GHz from
2003 through 2007. From the bottom to top, Stokes I, fractional linear polarization,
electric vector position angle, and fractional circular polarization. 
Two week averages of the data are shown.  The standard errors,
dominated by random measurement errors, are often smaller than the plotting symbols used.  
The horizontal line marks a zero level for circular polarization, and the vertical 
line indicates the epoch of the VLBA measurements.  
}
\end{figure}

The integrated total flux density, linear polarization, and circular 
polarization of 3C 279 have been monitored with the University of Michigan 
26 meter telescope operating alternately at 4.8, 8.0 and 14.5 GHz. The prime-focus
polarimeters, utilizing rotating quarter-wave plates feeding orthogonal 
linearly-polarized transducers, measure all four Stokes parameters simultaneously. The general observing 
and calibration procedures used are described in \citet{AAP03}. Each series of on-off
polarization observations, lasting approximately 40 minutes, was preceded by position 
scans to verify the pointing of the telescope. Observations of 3C 279 were interleaved 
with observations of reference sources at roughly two-hour intervals. The instrumental 
polarization was checked from observations of bright galactic HII regions which are 
assumed to be unpolarized.  All observations are restricted to within three hours of the meridian 
to minimize possible instrumental effects at large hour angles. Sources are not observed within 
15 degrees of the sun at 14.5 or 8.0 GHz or within 30 degrees of the sun at 4.8 GHz to avoid 
solar interference. Because of these restrictions, there are annual gaps in the data for 3C 279 
during September through mid-November at 4.8 GHz and from late September through late October 
at the two higher frequencies.  Figure \ref{f:UMRAO} shows the integrated emission of 3C\,279 
from 2003 through 2007.
The annual gaps in the data are due to the close proximity of the sun at those times.  We detect 
non-zero circular polarization at all three observing frequencies, and within the measurement 
uncertainties the Michigan instrument measured the same amplitude and polarity of circular 
polarization as observed by the VLBA.  During the period shown, circular polarization  
exhibited a preference to be negative at 4.8 GHz and positive at the two higher frequencies.

\vspace{0.3in}
\section{Deriving Physical Parameters of the Emission Regions}
\label{s:rad}

Our objective in this section is to model the magnetic field structure 
and particle properties in the core region of the parsec-scale jet of 3C\,279.
Tables 1 and 2 provide a large number of 
constraints: Stokes $I$ spectra for each component, fractional linear 
polarization (LP), $m_L$, as a function of frequency for each component, electric
vector position angle (EVPA), $\chi$, as a function of frequency for each component,
and finally, fractional circular polarization (CP), $m_c$, for the entire core region
at low frequency and for component ``D'' at high frequency.  

In the analysis that follows, we take the jet to have a bulk Lorentz factor 
of $\Gamma = 15$ at an angle of $\theta = 1.5^\circ$ to the line of sight,
giving a Doppler factor of $\delta = 26$ for the jet.  These numbers are
consistent with the recent kinematical analysis of \citet{H03}.  Similar 
but slightly larger values are found by \citet{J04}: $\Gamma = 20$ and 
$\theta=1.0^\circ$, giving $\delta = 36$.  Both sets of values give an 
observed angle in the 
frame of the emitting fluid of $\theta' \simeq 40^\circ$, and our results 
are not strongly dependent on which set of values we choose.  If we did choose
the ``larger'' value for the Doppler factor, it becomes somewhat easier 
to produce circular polarization via the intrinsic mechanism, as the 
estimated field strengths derived for the emission regions are somewhat 
larger (see \S{\ref{s:specI}}). 

\subsection{Major Models of CP Production}
\label{s:models}

Before we begin the detailed analysis, we will briefly outline five 
major models of circular polarization production which may be relevant
to this source \citep[e.g.][and references therein]{WH03}.

Model (1) is stochastic production of circular polarization from a 
purely tangled magnetic 
field.  In this 
model, the CP could be produced by either the intrinsic
or Faraday conversion mechanisms, or some combination of the two.
This model requires a fairly coarse-grained magnetic field tangling
to do the job, as the sign of the produced circular polarization will 
differ from cell-to-cell, leading to square-root$-$N style cancellation.  A
key prediction of this model is that the circular polarization should
vary in sign and amount across different frequencies, which probe 
different parts of the jet near optical depth equals unity.  We would
also expect in this model that the CP would vary with time as the
details of the magnetic field tangling change as the jet flows outward.
Our observations of 3C\,279 clearly show a consistency of sign and magnitude
of circular polarization across the frequency bands, and previous work 
\citep[e.g.][]{HL06} has shown that 3C\,279 has maintained the same sign
of circular polarization at high frequency for a decade now despite having a
fast flowing jet and several outbursts over that period.  The UMRAO
has also observed a consistent sign of positive CP in 3C\,279 at 14.5 GHz
from 2003 through 2007, see figure \ref{f:UMRAO}; however, the UMRAO also finds periodic 
switches to negative CP at 4.8 GHz which we discuss in \S{\ref{s:discuss}}
as a possible opacity effect.  Based on these results, we conclude 
that stochastic production of CP in a random magnetic field
is not a major mechanism producing the CP we see in 3C\,279.  In 
the analysis that follows, we will therefore assume that any tangled portion of
the magnetic field in 3C\,279 is tangled on a very short length-scale
to minimize the production of stochastic CP and allow exploration
of the contribution by organized magnetic fields. 

Model (2) is intrinsic circular polarization from a strong, 
vector-ordered magnetic field, hereafter {\em intrinsic circular
polarization}.  Given the consistency across epoch
and frequency in the observed circular polarization of 3C\,279, the most
plausible candidate for such a vector-ordered field in the jet 
is a poloidal field along the jet axis.  We can estimate the 
strength of such a field.  If the magnetic field were completely uniform 
(no reversals or tangling), the expected circular polarization
would be of order $m_c \sim \sqrt{\nu_B/\nu_{emit}}$ where $\nu_B = 2.8B$ 
MHz is the electron gyro-frequency and $\nu_{emit} = \nu_{obs}(1+z)/\delta$ 
is the emitted frequency \citep[e.g.,][]{WH03}.  So for $1\%$ CP at 
22 GHz, the required magnetic field strength is about $B\sim 50$ mG.
This is not an unreasonable field strength, and as we shall see in \S{3.4.2}, is 
of the order of the estimated field strength in component ``D''.  

It is important to note that the above estimate assumes not only a 
uniform magnetic field, but also that all of the emitting particles are 
electrons.  This will be the case in a ``normal'' matter jet consisting
of a pure electron-proton plasma.  Allowing for the possibility of some
admixture of electron-positron pairs, parameterized by the lepton 
number, $\ell$, defined in the introduction, the fractional circular polarization is
$m_c \propto \ell$ and the required field strength would 
scale as $B \propto \ell^{-2}$.

Model (3) is Faraday conversion of linear polarization
into circular polarization driven by Faraday rotation, hereafter 
{\em rotation driven conversion}.  In this model the LP is produced 
by whatever magnetic field 
order is available in the jet, such as shocked or sheared magnetic 
field.  However, that LP cannot be directly converted into CP by the 
same field order because there needs to be some offset between the 
position angle of the LP and the angle of the magnetic field 
doing the conversion.  The angular offset is provided by internal
Faraday rotation within the jet, and in this model, the Faraday rotation 
depth is relatively small so that the field at the front of the jet is 
converting the LP emitted by the back of the jet into some CP. 
 
For this to be a consistent (not stochastic) process, the Faraday 
rotation must be provided by some vector-ordered field in the jet,
probably a poloidal field along the jet axis as in model (2).  Additionally, there must
also be some preponderance of electrons in the jet relative to 
positrons, i.e., $\ell > 0$.  However, the requirements for some
vector-ordered field and some preponderance of electrons in the jet
plasma are not nearly as strong as they are for model (2), where the
CP is produced entirely by the intrinsic mechanism.  Model (3) has the
additional requirement of low energy particles within the jet to 
produce the internal Faraday rotation, either 
due to a low cutoff, $\gamma_l$, in the relativistic particle power-law 
spectrum $N_\gamma d\gamma = K\gamma^{-p}d\gamma$ for $\gamma_l \leq 
\gamma \leq \gamma_u$ or due to the addition of some ``cold'', 
non-relativistic thermal matter to the jet.

It is interesting to note that the similarity in requirements between
models (2) and (3) means that they will often act together to at least 
some degree.

Model (4) is a high rotation depth version of model (3), proposed and
investigated independently by \citet{RB02} and \citet{BF02}.  The 
requirements are almost the same as for model (3), but the rotation
depth is much larger, so large, in fact, that significant rotation and
conversion can happen over very small length scales in the jet.  As long
as this sense of rotation is common in all parts of the jet, no large scale
magnetic field order is required (except for the vector-ordered field 
which produces the rotation) and net circular polarization can be 
produced in potentially large amounts with little if any net linear 
polarization \citep{BF02, RB02}.

Model (5) is Faraday conversion from a helical magnetic field or other
field structure which varies systematically in orientation across the jet.  In
this model, no internal Faraday rotation is required, as the field orientation
at the back of the jet is already at some angle with respect to the field
orientation at the front of the jet \citep[e.g.][]{WH03, G08}. 

\subsection{Numerical Modeling}
\label{s:rad_model}

We have written a numerical simulation that solves the full-Stokes equations
of radiative transfer \citep{JOD77,Jones88} either by numerical integration
or, optionally, by using the exact solutions calculated by \citet{JOD77}.
An early version of this simulation was used by \citet{WHOR98} to interpret
the first parsec-scale circular polarization observations on 3C\,279.  The
simulation models the emission from a ``homogeneous'' line of sight broken
into cells where each cell has the same physical properties, including 
spectral index, $\alpha$, lepton number, $\ell$, Doppler factor, $\delta$, 
viewing angle, $\theta$, and low energy power-law cutoff for the relativistic 
particle distribution, $\gamma_l$.   The exception to this is the 
magnetic field, which may consist of multiple components which can vary in 
direction and/or 
magnitude from cell to cell depending on the magnetic field model applied.
Additionally, multiple lines of sight can be put together to construct simple 
inhomogeneous 
models (see \S{\ref{s:specP_compD}}) or homogeneous models where 
the magnetic field has transverse
structure, such as toroidal or helical magnetic fields 
(see \S{\ref{s:specP_comp54}}).  
By adjusting the physical parameters and magnetic field model applied, we can 
simulate any of the five conceptual models described in the previous 
section.  The radiative transfer is done in the frame co-moving with the fluid, and here 
we have assumed a single flow velocity where the relativistic flow parameters are
derived from observed pattern speeds for this jet given by VLBA kinematical analyses 
as described in the introduction to this section.  This simulation is static in the sense
that the magnetic field and particle properties of the emission region are assumed not 
to change during a light crossing time.

\subsubsection{Parameterization of the Magnetic Field}
\label{s:bfield_par}

The magnetic field in our simulation can have three different components: 
a vector-ordered uniform field along the jet axis, $B_u = B_\star\times f_u$,
a toroidal field, $B_t = B_\star\times f_t \times (\rho/\rho_{jet})$, and 
a randomly ordered field which varies stochastically from cell to cell, 
$B_r = B_\star \times (1 - f_u - f_t)$.  Here $B_\star$ is a scaling factor
which allows us to match a desired average perpendicular magnetic field 
strength, $<B_\perp>$, set by observation,  $f_u$ and $f_t$ parameterize
the degree of uniform and toroidal fields respectively, and 
$\rho/\rho_{jet}$ is the fractional distance of a cell from the center axis
of the jet, where $\rho/\rho_{jet} = 1.0$ is a cell at the outside edge 
of the jet.  This parameterization of the toroidal field assumes that the 
current carried by the jet is uniformly distributed.  Note that non-zero values
for $f_u$ and $f_t$ together will produce a helical field order.

In addition to the three field components described above, the magnetic
field may also be shocked, by shortening unit length to length, $k \leq 1.0$.
This shock is assumed to be a transverse shock, so the shortening occurs
along the jet axis.  The result is amplification of the magnetic field
components transverse to the jet axis as described in detail in the 
appendix of \citet{WCRB94}.

\subsection{Modeling the Stokes $I$ Spectra}
\label{s:specI}

\begin{figure}
\figurenum{6}
\includegraphics[scale=0.37,angle=-90]{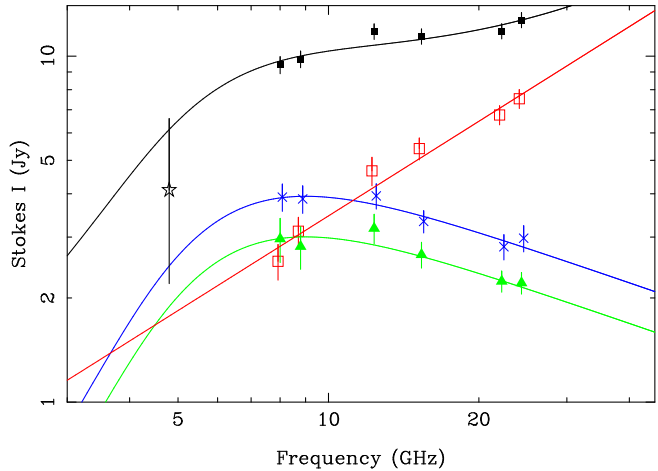}
\figcaption[]{\label{f:specI}
Stokes $I$ spectrum of the core region of 3C\,279.  Component D is plotted
with red open squares.  Components 5 and 4 are plotted with blue $\times$
symbols and green filled triangles respectively.  The combined core region
flux is plotted in black filled squares.  An estimate for the combined 
core region flux at 4.8 GHz from comparison of the UMRAO and VLBA data
is plotted as an open star.  The spectral fits described in \S{\ref{s:specI}}
are plotted as solid lines of the same color as the corresponding components.
}
\end{figure}

Figure \ref{f:specI} shows the Stokes $I$ spectra of components D, 5, and 4 and
their total flux density along with analytical models for each component.  The total
core flux at 4.8 GHz is estimated by comparing quasi-simultaneous UMRAO integrated 
measurements at 4.8, 8.0, and 14.5 GHz to our VLBA measurements.

Components 5 and 4 are modeled analytically as homogeneous synchrotron sources, but
when we fit their six-frequency spectra using the Stokes $I$ data
alone, we found that the spectral turnover frequencies, $\nu_{peak}$ were not
well constrained at low frequency (see the first line in Table 3 for 
both components 5 and 4).  However, the fractional linear polarization 
of both components decreased with frequency in a manner consistent
with turnover frequencies closer to $9.0$ GHz for both components.
To quantify this observation we developed an analytical model of 
how fractional linear polarization, $m_l$, should decrease with 
increasing optical depth, including the possible effects of
depolarization from internal Faraday rotation.  This model is described
in detail in the Appendix.

\begin{figure}
\figurenum{7}
\includegraphics[scale=0.75,angle=0]{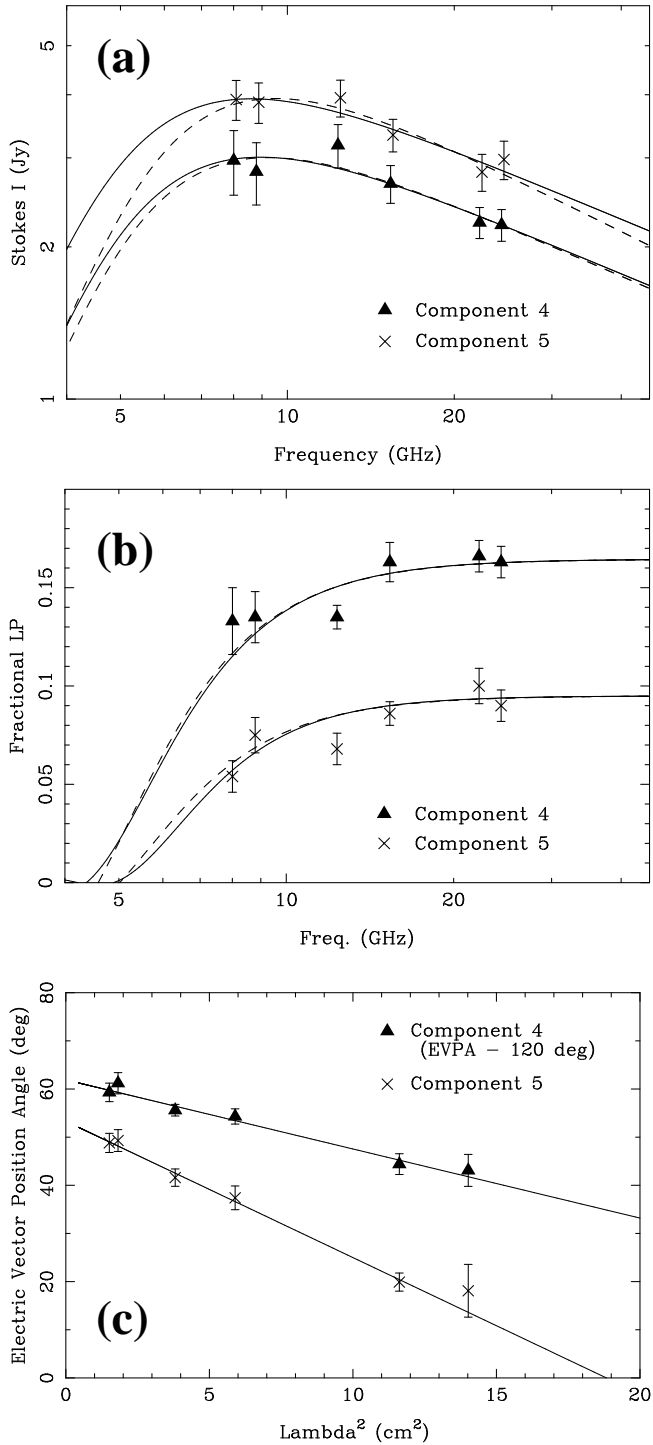}
\figcaption[]{\label{f:spec_comp54}
Stokes $I$ (panel (a)), fractional linear polarization (panel (b)), 
and electric vector position angles (panel (c)) of components 5 and 4 
plotted against two homogeneous spectrum models discussed in section 3.3.  
The dashed lines in panels (a) and (b) represent a joint fit to the Stokes $I$ and 
fractional linear polarization data assuming none of the observed 
Faraday rotation is internal to the jet.  The solid line in these panels 
includes a modified Burn-style depolarization in the fit assuming that 
all of the observed Faraday rotation is internal to the jet.  A $\lambda^2$ 
regression to the observed Faraday rotation is plotted in panel (c) as a 
solid line.  
}
\end{figure}

We repeated the analytical fits including the fractional linear
polarization data, both with and without the possibility of
depolarization being due to the observed Faraday rotation being
internal to the jet.  The results for these more highly constrained fits
are included in Table 3 and plotted in Figure \ref{f:spec_comp54}, 
panels (a) and (b).  To quantify
the degree of observed Faraday rotation in these components
we fit a regression of the standard form 
$\chi_{obs} = RM\lambda^2 + \chi_{emit}$ where we found 
$RM = -493\pm23$ ${\mathrm rad\;m^{-2}}$ and $RM = -250\pm25$ ${\mathrm rad\;m^{-2}}$ 
for components 5 and 4 respectively.  These regressions are plotted in
Figure \ref{f:spec_comp54}, panel (c).  
Because the total rotations in these two cases are 
relatively small, a $\lambda^2$ regression works well regardless of
whether the observed rotation is internal or external to the 
jet.  For both components 5 and 4, our results, including or excluding
the possibility of internal Faraday rotation, are consistent with
peak frequencies of $\nu_{peak} = 9.0\pm1.0$ GHz and spectral
indices of $\alpha = -0.5\pm0.1$ (we use the convention 
$S\propto \nu^{+\alpha}$).  Because our analysis is
not particularly sensitive to the precise location of the turnover, 
we adopt these values in our subsequent analyses (they are also
used to plot the spectra which appear in Figure \ref{f:specI}).

\begin{deluxetable}{ccrrr}
\tablecolumns{5}
\tablewidth{0pc}
\tabletypesize{\scriptsize}
\tablenum{3}
\tablecaption{Spectral Turnover Fits to Components 5 and 4.\label{t:spec_fits}}
\tablehead{\colhead{Component} & \colhead {Method} & \colhead{$\nu_{peak}$} & \colhead{$S_{peak}$} & 
\colhead{$\alpha$} \\
\colhead{ID} & & \colhead{(GHz)} & \colhead{(Jy)} &  \\
\colhead{(1)} & \colhead{(2)} & \colhead{(3)} & \colhead{(4)} &
\colhead{(5)}}
\startdata
  5 & $Stokes$ $I$ $Only$ & $8.95^{+1.83}_{-8.76}$ & $ 3.94^{+4.99}_{-0.16}$ & $-0.50^{+0.21}_{-0.20}$ \\ 
    & $Stokes$ $I$ $+$ $m_l$ & $9.51^{+0.63}_{-0.53}$ & $ 3.93^{+0.20}_{-0.18}$ & $-0.55^{+0.10}_{-0.09}$ \\ 
    & $Stokes$ $I$ $+$ $m_l$ $+$ $FR$ & $8.61^{+0.77}_{-0.89}$ & $ 3.93^{+0.24}_{-0.21}$ & $-0.46^{+0.10}_{-0.10}$ \\ 
  4 & $Stokes$ $I$ $Only$ & $10.21^{+1.11}_{-5.19}$ & $ 3.06^{+2.07}_{-0.09}$ & $-0.62^{+0.22}_{-0.22}$ \\ 
    & $Stokes$ $I$ $+$ $m_l$ & $9.23^{+0.86}_{-0.69}$ & $ 3.00^{+0.23}_{-0.19}$ & $-0.48^{+0.11}_{-0.10}$ \\ 
    & $Stokes$ $I$ $+$ $m_l$ $+$ $FR$ & $8.94^{+0.95}_{-0.69}$ & $ 3.01^{+0.24}_{-0.20}$ & $-0.46^{+0.11}_{-0.11}$ \\ 
\enddata
\tablecomments{
Columns are as follows: (1) Component identifier; 
(2) Method of fitting spectral peak, as described in \S{3.3}; 
(3) Observed peak frequency in GHz;
(4) Peak Stokes-I flux density in Jy;
(5) Spectral index. 
}
\end{deluxetable}

Component D has an inverted spectrum and is not well fit by a homogeneous 
source spectrum.  It can be most simply fit by a straight power-law with 
spectral index $\alpha = +0.91\pm0.08$.  
In \S{\ref{s:specP_compD}} we will model component D as an inhomogeneous conical 
jet with power-law dependence on magnetic field strength and particle density
as described by \citet{BK79} and \citet{K81}.

\subsubsection{Estimating Magnetic Field Strength}
\label{s:bfield_strength}

We can estimate the magnetic field strengths in components 5 and 4 by
treating them as homogeneous spheres \citep[e.g.][]{M87}.  For a
homogeneous volume of plasma \citep[][eqn. A1]{HW00}

\begin{equation}
B = \frac{\delta}{(1+z)}C_3\tau_m^2\nu_m^5S_m^{-2}\Omega^2
\end{equation}

Where $C_3 = 5.30\times 10^{-4}$ for $\alpha = -0.5$, $\Omega$ in mas$^2$, 
$\nu$ in GHz, $S$ in Jy, and $B$ in Gauss.  For an homogeneous
sphere at the turnover frequency with $\alpha = -0.5$, 
$\tau_m = 0.48$, $S_m$ is 1.19 times the observed flux $S_o$, and
$\Omega = (\pi/6)\theta^2_d$ where $\theta_d$ is the angular diameter
of the component \citep{HW00}.  We use the approximation, 
$\theta_d \simeq 1.8 (\theta_a \theta_b)^{-1/2}$, to convert Gaussian
fitted FWHM dimensions to spherical diameters \citep{M87}.

Taking the median angular sizes fitted amongst our models, we
estimate $B \sim 0.6$ mG for component 5 and $B \sim 4.4$ mG 
for component 4.  These estimates depend on the assumed
geometry and are probably only good to a factor of a few; 
however, our results are not strongly dependent on these 
values.   

For the inhomogeneous component D, we can apply the above 
technique to make an order of magnitude estimate for its magnetic
field strength; however, in that case, it would be more appropriate 
to take an optical depth $\tau$ between 1 and 2, and we estimate 
$\tau \sim 1.5$.  Using the observed values at 22 GHz, we 
estimate $B \sim 40$ $mG$.  In the following section we will 
allow for a range of possible magnetic field strengths for 
component D.

\subsection{Modeling the Polarization Spectra}
\label{s:specP}

\subsubsection{Components 5 and 4}
\label{s:specP_comp54}
The Stokes $I$ spectra of components 5 and 4 are adequately
modeled by homogeneous components.  Both features show strong
linear polarization and modest amounts of Faraday rotation: 
$\simeq -35^\circ$ to $-40^\circ$ and $\simeq -15^\circ$ to 
$-20^\circ$ of total rotation by 
8.0 GHz for components 5 and 4 respectively.  We have no direct
evidence for how much if any of this observed rotation is internal 
to the jet as neither of these rotations are enough 
to generate large amounts of internal depolarization.  In
the following analysis, we will construct two plausible models,
one where all of the observed rotation is internal and the other 
where none of the rotation is internal to the jet.

Our observations indicate that very little, if any, of the observed 
circular polarization at 22 and 24 GHz is produced in these
two features; however, by 8.0 and 8.8 GHz we cannot separate the
circular polarization between D, 5, and 4.  Indeed, if we
assign all the circular polarization observed at 8.0 and 8.8 GHz
to component D alone, we get very large levels of $1.99\pm0.38$\%
and $1.55\pm0.26$\% respectively.  These amounts are far larger
than the amounts observed on D at 15, 22, and 24 GHz, where we find
less than $1$\%.  So, either component D must produce a sharp rise 
in circular polarization at these two frequencies, or components 
5 and 4 make a significant contribution to the total circular 
polarization at 8.0 and 8.8 GHz but not at 15 GHz and above.  
To accomplish this second scenario, components 5 and 4 must each 
produce of order $0.5-1.0$\% circular polarization at 8.0 GHz while 
producing $\lesssim 0.2$\% at 22 and 24 GHz.

In reference to the major models of circular polarization production 
considered in \S{\ref{s:models}}, only models (3) and (5) are
plausible for these components.  Model (2), intrinsic circular
polarization, needs stronger magnetic field strengths and a higher
degree of magnetic field order.  Additionally, the spectral 
dependence of intrinsic CP, $m_c \propto \nu^{-0.5}$, is not
steep enough.  Model (4) requires very large rotation depths 
which are not observed here.  

Model (3) from \S{\ref{s:models}} is Faraday conversion driven
by internal Faraday rotation, so for this model, we will assume
that all of the observed rotation is internal to the jet.  In
this model, the magnetic field order which creates the linear
polarization at the back of the jet is the same field order 
which does the conversion to circular polarization at the
front of the jet. Faraday rotation within the jet rotates
the linear polarization by angle $\phi$ to allow conversion 
by this same field, where the circular polarization produced
is proportional to $\sin 2\phi$.  So for this model to 
operate effectively, we need two components of organized 
magnetic field.  The first component is in the plane
of the sky and first produces the linear polarization and 
later converts it into circular polarization.  The second
component is vector-ordered magnetic field which produces
the internal Faraday rotation.

For this model we will assume the first component of magnetic
field is provided by a transverse shock.  
In component 5, the unrotated electric vector position angle 
(EVPA) is approximately $54^\circ$, which gives a magnetic field 
position angle of $-36^\circ$, nearly perpendicular to the 
structural position angle of $-121^\circ$ for component 5, indeed suggesting that 
the dominant field order is due to a transverse shock.  
For component 4, the unrotated EVPA is approximately $2^\circ$,
giving a magnetic field position angle of $-88^\circ$ which is neither
perpendicular nor parallel to its structural position 
angle of $-143^\circ$, suggesting that the field order in 
component 4 is due to an oblique shock.  Our radiative transfer
simulation is set-up to allow transverse shocks, see 
\S{\ref{s:bfield_par}}; however, for the purposes of simulating
the effects of Faraday conversion in this model, the differences 
between an oblique and transverse shock are unimportant, as we
only need the shock to provide field order in the plane of the 
sky to first produce the linear polarization and then to convert 
it into circular polarization.

\begin{figure}
\figurenum{8}
\includegraphics[scale=0.9,angle=0]{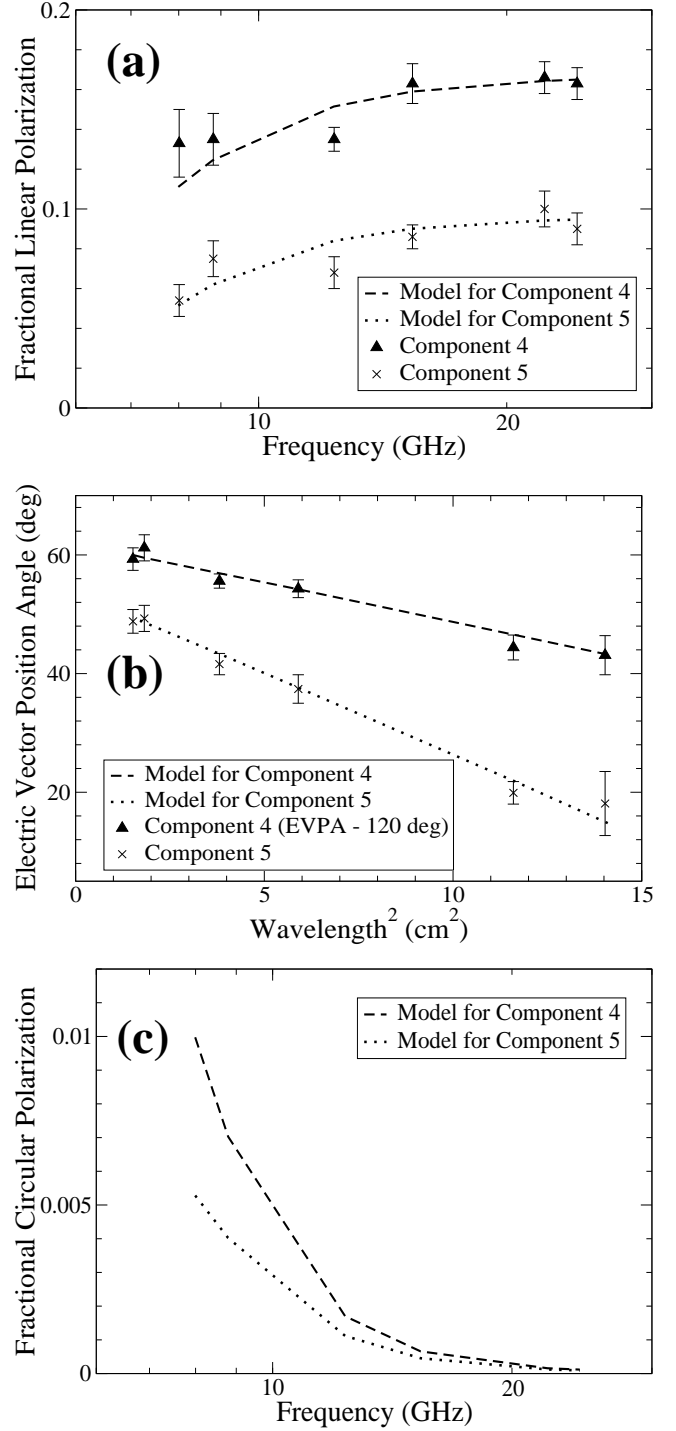}
\figcaption[]{\label{f:spec_comp54_pol}
Linear and circular polarization spectra for components 5 and 4
from the rotation driven conversion model described in 
\S{\ref{s:specP}}.  Panel (a) plots the fractional linear 
polarization model against the data, panel (b) plots the 
electric vector position angle (EVPA) against the data, and 
panel (c) plots the predicted fractional circular polarization 
as a function of frequency.  The model EVPA values in panel (b) have
been shifted by a constant offset to align with the high 
frequency data.  Note that in panel (b), the EVPA for component 
4 has been shifted by $-120$ degrees to allow it to fit comfortably on this plot.
}
\end{figure}

For the second component of magnetic field which generates the
internal Faraday rotation, we allow a vector-ordered field along 
the jet axis parameterized as $f_u = 0.05$ as described in 
\S{\ref{s:bfield_par}} (note that $f_t = 0$ here so there is no
contribution from a toroidal field component).  With this degree 
of vector-ordered field along the jet axis, the shocks in 
components 5 and 4 must be of strength $k = 0.64$ and $k = 0.27$ 
respectively to give the observed amounts of linear polarization at high 
frequency.  We then put in the magnetic field strengths estimated
in section \S{\ref{s:bfield_strength}}, and adjust the lower
cutoff in the relativistic particle spectrum, $\gamma_l$, to give
the observed amount of Faraday rotation: $\gamma_l = 19$ and 
$\gamma_l = 8.5$ for components 5 and 4 respectively.  The optical
depth at each frequency is set by the Stokes $I$ spectra determined
in \S{\ref{s:specI}}.  For each component, we assume a lepton
number of $\ell = 1.0$, representing a pure electron-proton plasma.
With these parameters set, we compute the radiative transfer along
a line of sight with $N = 10^6$ cells to minimize 
stochastic effects from the random components of the magnetic 
field.  The results of this calculation appear in Figure 
\ref{f:spec_comp54_pol}.

This model produces the correct sign, amount, and 
spectrum of circular polarization that we need for 
components 5 and 4, with $m_c$ in the range 
of $+0.5$ to $+1.0$\% at 8.0 GHz and
falling off to negligible values at high frequency.  An
important question is:  how sensitive are these values
to the parameters chosen in the paragraph above?  In short,
these values are sensitive only to one parameter: the
amount of internal Faraday rotation.  In the above model
we have assumed that all of the observed Faraday rotation is
internal to the jet.  For modest optical depths, the
circular polarization produced by Faraday conversion 
driven by Faraday rotation is proportional to the degree
of magnetic field order squared, which is fixed by the 
observations of the linear polarization, and to the
product $\tau_F\tau_C$, where $\tau_F = \zeta^*_V\tau$ and 
$\tau_C = \zeta^*_Q\tau$ are the Faraday rotation and conversion 
depths respectively \citep{WH03, JOD77}.  The optical
depth, $\tau$, is also fixed by observation, so it is only
the coefficients which matter.  For $\alpha = 0.5$, the 
coefficients are given by

\begin{equation}
\zeta^*_V \simeq \zeta^{*V}_\alpha \ell f_u \frac{\nu}{\nu_{B\perp}}
               \frac{\ln\gamma_l}{\gamma_l^3} \cot\theta'
\end{equation}

and

\begin{equation}
\zeta^*_Q = 2 \zeta^{*Q}_\alpha \ln\frac{\gamma}{\gamma_l}
\end{equation}

where $\gamma = \sqrt{\nu/\nu_{B\perp}}$ is the $\gamma$ of the
radiating particles, and $\xi^{*Q}_\alpha$ and $\xi^{*V}_\alpha$ are
constants of order unity.  Note that the dependence on $f_u$ given
for $\xi^*_V$ is approximate and only good for small values of 
$f_u \lesssim 0.1$.

In summary, if all of the observed Faraday rotation is 
assumed to be internal, then $\tau_F$ is also fixed by
observation and the combination of parameters 
$\ell f_u B^{-1}_{\perp}\ln\gamma_l / \gamma_l^3$ is likewise
fixed.  Thus, the only place where parameter choice 
affects the results of this model is through the conversion
depth $\tau_C$ which depends only weakly on the magnetic
field strength, $B_\perp$ and the lower cutoff, $\gamma_l$,
through the logarithm $\ln{(\gamma/\gamma_l)}$. In
this respect, the rotation driven conversion model is
attractive as the sign, amount, and spectrum of circular 
polarization produced is nearly fixed by the observed 
properties of the linear polarization.

In contrast, the helical magnetic field model, model (5)
from \S{\ref{s:models}}, has fewer constraints because here
we assume that all of the observed Faraday rotation
is external to the jet, so as to assess the ability of the 
helical field itself to produce the observed circular
polarization.  As above, we assume $f_u = 0.05$ for
the vector-ordered component of magnetic field along
the jet axis, and with this value, we require a 
toroidal field of $f_t = 0.56$ and $f_t = 0.78$ to
match the observed levels of linear polarization in
components 5 and 4 respectively.  We note that the
unrotated EVPA of component 4, computed above, is
not consistent with a toroidal field which should be
perpendicular to the jet axis; however, for the 
purpose of this analysis we will ignore this 
inconsistency.

\begin{figure}
\figurenum{9}
\includegraphics[scale=0.9,angle=0]{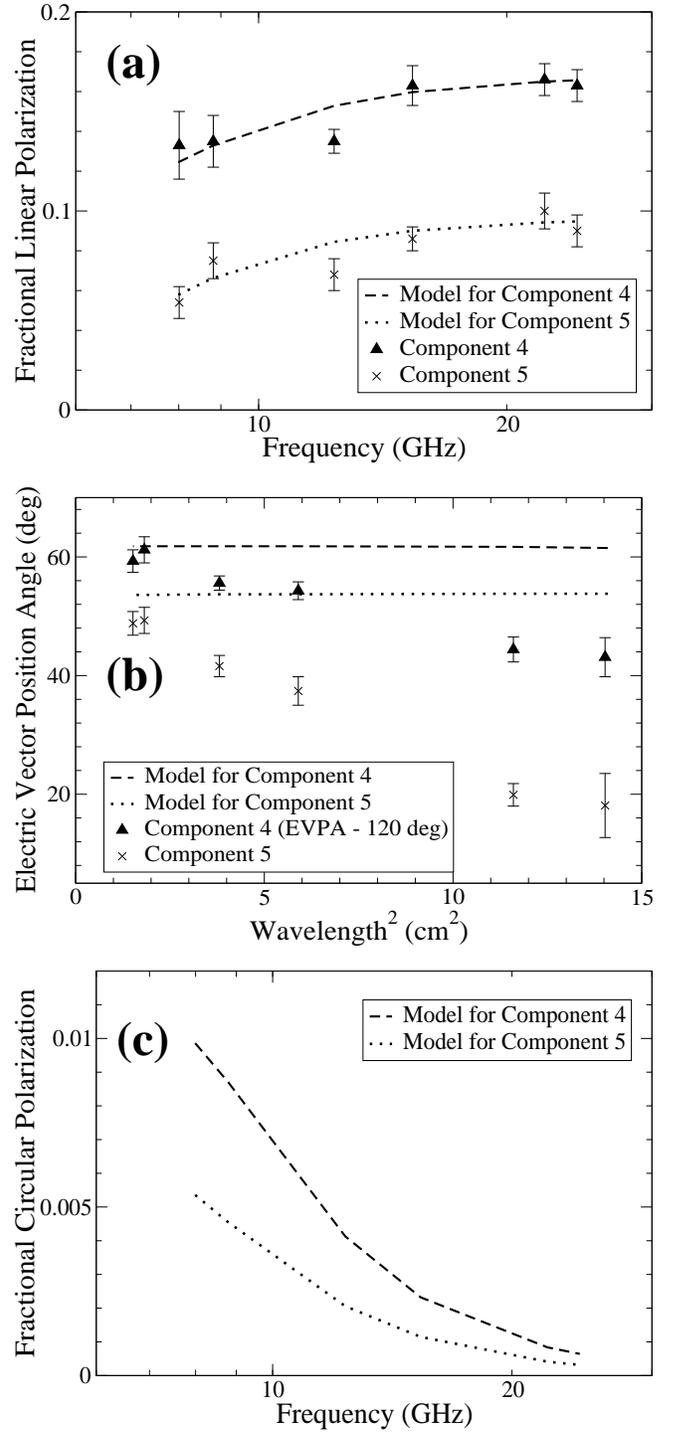}
\figcaption[]{\label{f:spec_comp54_hel}
Linear and circular polarization spectra for components 5 and 4
from the helical field conversion model described in 
\S{\ref{s:specP}}.  Panel (a) plots the fractional linear 
polarization model against the data, panel (b) plots the 
electric vector position angle (EVPA) against the data, and 
panel (c) plots the predicted fractional circular polarization 
as a function of frequency.  The model EVPA values in panel (b)
have been shifted by the same constant offset used in Figure 
\ref{f:spec_comp54_pol}. Note that in panel (b), the helical
field model predicts a flat EVPA with frequency because none
of the observed Faraday rotation is assumed to be internal to
the jet.  As in Figure \ref{f:spec_comp54_pol}, the EVPA for 
component 4 has been shifted by $-120$ degrees to allow it to 
fit comfortably on this plot.
}
\end{figure}

With $f_t \gg f_u$, the overall field-order is a 
high pitch-angle helix as suggested by \citet{G08} 
for 3C\,279, and we can obtain similar levels of
circular polarization as the rotation driven 
conversion model by choosing $\gamma_l = 20$ and
$\gamma_l = 10$ for components 5 and 4 respectively.
We have also taken $\ell = 10^{-5}$ to completely
eliminate any internal Faraday rotation, and no
shock has been assumed for either component, $k = 1.0$.
For a helical field, where the field order varies across
the jet cross-section, a single line of sight is not
sufficient to determine the emergent polarization, so we
take a $100\times100\times100$ cube and set to zero any 
cells outside a cylindrical jet cross-section.  The
emerging polarization for both components is plotted in 
Figure \ref{f:spec_comp54_hel}.

As with the rotation driven conversion model, the
helical field is able to produce circular polarization
with the correct amplitude, sign, and spectra; however,
the amount and sign of the produced circular polarization
are no longer determined by the linear polarization and
are the result of choices we made about field direction
and the magnitude of $\gamma_l$.  The spectrum of the 
circular polarization from the helical field falls off 
more slowly than for the rotation driven conversion 
case, but the levels are small enough at 15 GHz and
above to still be acceptable.  This shallower spectrum
is due to the fact that the helical field produces CP 
via pure Faraday conversion and depends only upon $\tau_C$,
whereas rotation driven conversion depends on the product
of $\tau_F\tau_C$ which includes two powers of the 
optical depth.

\subsubsection{Component D} 
\label{s:specP_compD}

Component D represents the base of the jet as observed in
our VLBA images, and it has an inhomogeneous spectrum 
which is well fit as a single power-law $S\propto\nu^{+\alpha}$ 
where $\alpha = +0.91\pm0.08$ as described in \S{\ref{s:specI}}.
Its linear polarization is approximately flat or perhaps
slightly decreasing with frequency from 
$m_l \simeq 3.6$ to $2.3$\% from low to high frequency.
The EVPA of the linear polarization stays roughly 
constant between $+110^\circ$ to $+120^\circ$ except for
24 GHz, where we find $+101^\circ\pm3^\circ$.  Its circular
polarization at 22 and 24 GHz is $0.86\pm0.20$\% and 
$0.94\pm0.22$\% respectively, and if we assign all the observed
circular polarization at 15 GHz to component D, we find a
similar amount of $0.88\pm0.15$\%.

These properties are all suggestive of an inhomogeneous
component of the type investigated by \citet{BK79} and
\citet{K81} where the magnetic field and particle densities
scale as power laws in a conical jet with $B_\perp \propto r^{-m}$
and $K \propto r^{-n}$.  The EVPA of the linear polarization
described above corresponds to a jet magnetic field which is neither
perpendicular nor parallel to the jet axis. The later is assumed to
lie somewhere between the structural position angles 
of components 5 and 4 of $-121^\circ$ and $-143^\circ$
respectively.  Additionally, the observed levels of 
linear polarization at high frequency (2$-$3\%) are 
too low to generate the nearly $1$\% corresponding circular 
polarization if the linear polarization is taken as a 
direct measure of the field order.  In all likelihood,
Faraday rotation and depolarization, either internal or external
to the jet, are responsible for the offset EVPAs and the
low observed levels of linear polarization.  If the culprit Faraday
screen is external to the jet, then it must scale
in a similar power-law fashion in order for the jet properties themselves
to maintain an approximately constant level of depolarization
and rotation with frequency.  While such a scaling of an external
screen is certainly possible, it seems more likely that the 
depolarization and rotation is occurring internal to the jet, 
where it naturally would scale in the appropriate manner with 
$B$ and $K$.  

In constructing our inhomogeneous model of component D, we
will assume that all of the observed depolarization and 
rotation are occurring internal to the jet, matching our
assumption for the rotation-driven conversion model of
components 5 and 4.  This assumption provides the most
stringent constraints on component D, as the same magnetic
field model must produce the correct signs and amounts of
circular polarization, Faraday rotation, and depolarization.
From \S{\ref{s:models}}, models (2), (3) and (4) for circular
polarization production all fit this scenario, and we can 
simultaneously investigate all three by constructing 
inhomogeneous jets consisting of vector-ordered field 
along the axis and disordered field which may be shocked
from unit length to length k as described in 
\S{\ref{s:bfield_par}}.  

We simulate conical, inhomogeneous jet emission by running our
radiative transfer simulation for many lines of sight
through the center of a jet where each line of sight 
corresponds to a new radius, r, from the base of the
jet.  The magnetic field strength and frequency where
$\tau = 1$ are scaled at each radius, r, in a power-law
fashion given by \citet{K81} where $B_\perp \propto r^{-m}$
and $\nu \propto r^{-k_m}$, where $k_m$ is given by \citet{K81}
and corresponds to a power-law scaling of not only $B_\perp$ but
also the particle density, $K\propto r^{-n}$.  In all of our
models, we assume an optically thin spectral index of 
$\alpha = -0.5$.  For $m = 1.8$ and $n = 2.8$, $k_m = 1.8$ and
we find a good match to the Stokes $I$ spectrum from our model.
For these values, \citet{K81} would predict $S \propto \nu^{+0.89}$,
which agrees well with our simulation.  Other combinations
of $m$ and $n$ could also match our Stokes $I$ spectrum, and we
explore three such combinations.

We set the magnetic field strength and degree of vector-ordered
field at an observed frequency of 22 GHz.  We require the vector
ordered field to scale like $r^{-2}$ to conserve magnetic flux, and
we adjust the scaling of the random component of the field to give
the overall correct scaling for $B_\perp \propto r^{-m}$.  For each
location in the jet we compute the emerging radiation at 
our VLBA observing frequencies and integrate the results from the 
entire inhomogeneous component, cutoff at radii which produce low 
and high frequencies of $1$ GHz and $100$ GHz respectively in the 
observer frame.  The Stokes $I$ spectrum is then scaled by a single 
multiplicative factor, common to all frequencies, to best align 
with our observations.  The model EVPA is also rotated by the same
angle at all frequencies to best align with the observed EVPAs 
between 15 and 24 GHz; however, this angle is not arbitrary as 
it tells us the direction of the net magnetic field in the jet of 
3C279 and should correspond to 
a sensible value for the magnetic field model of the simulation.  
The final result is a model spectra of Stokes $I$, fractional
linear polarization, EVPA, and circular polarization. 

We initially ran a coarse grid of models with $m=1.8$ and $n=2.8$,
exhausting all combinations of the following parameter values: 
$\ell = 1.0,0.1,0.01$, $f_u = 0.1$ to $0.9$ in $0.1$ steps, 
$B_\perp = 0.1, 0.05, 0.02, 0.01$ G, $\gamma_i = 3,5,8,10,15,25,50$,
and $k = 0.1$ to $1.0$ in $0.1$ steps.  We judged degree of agreement
between model and data by comparing the fractional circular 
polarization, fractional linear polarization, and EVPA at 15, 22,
and 24 GHz.  We don't expect every variation and wiggle in the 
observed data to be reproduced by the inhomogeneous power-law 
models.  In fact, we expect the inhomogeneous models to produce
smoother-than-observed values for the polarization; however, we
should be able to reproduce the general trends and levels observed.
In this light, we required ``plausible'' models to produce 
an average circular polarization in the range $m_c = 0.8$ to $1.0$\%, 
with no value smaller than $0.6$\% or larger than $1.3$\%, and
we required average linear polarization in the range $m_l = 2.2$ to $2.5$\%, 
with no value smaller than $2.0$\% or larger than $2.8$\%.  These
restrictions will only allow models to be plausible if they reproduce
the general levels and slopes in the observed data.  Out of more
than 7000 models in this coarse grid, we found only two that fit
these criteria; both models had $\ell = 1.0$, $B_\perp = 0.1$ G,
$f_u = 0.7-0.8$, $\gamma_l = 5$, and $k = 0.8-0.9$. 

\begin{figure*}
\figurenum{10}
\begin{center}
\includegraphics[scale=0.85,angle=0]{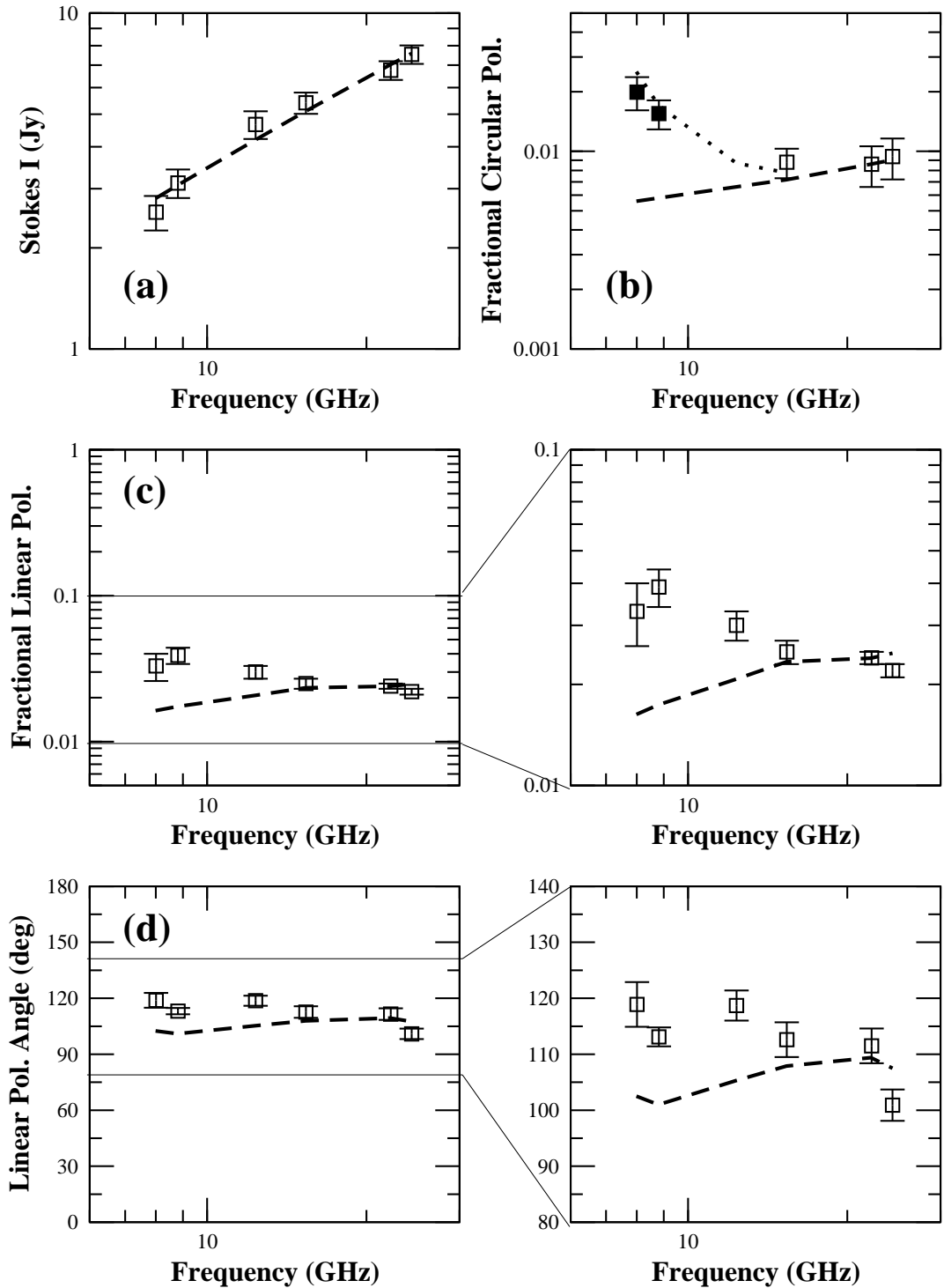}
\end{center}
\figcaption[]{\label{f:spec_compD_mod1cd}
Full polarization spectra of the inhomogeneous component D.  The 
data are plotted as open squares against a dashed line representing 
our best model described in \S{\ref{s:specP_compD}}. As described in
that section the quality of the agreement was judged only at 15, 22, and
24 GHz; however, we plot the full spectrum here.  Panel (a) is Stokes $I$.  
Panel (b) is fractional circular polarization.  In panel (b), the 
8.0 and 8.8 GHz values are plotted as solid squares to indicate that they
may include significant contributions from components 5 and 4.  The
results from our rotation driven conversion model in figure 
\ref{f:spec_comp54_pol} are added to the inhomogeneous component D model
and plotted as a dotted line in panel (b).  Panel (c) is fractional 
linear polarization with a zoomed-in panel to the right to show 
agreement more clearly in the region of interest.  Panel (d) is the 
electric vector position angle with a zoomed-in panel to more clearly 
show agreement in the region of interest.
}
\end{figure*}

Physically, these plausible models represent a jet dominated by
vector-ordered magnetic flux along the jet axis, producing large
amounts of internal Faraday rotation/depolarization with a 
significant contribution from intrinsic circular polarization
as well as rotation-driven conversion.  A range of similar models,
while not meeting our plausible criteria did show similar trends
in $m_l$ and $m_c$, and we wished to explore this type of  
physical model with a finer grid. This new grid looked at every 
combination of 
$\ell = 1.0, 0.5, 0.1$, $B_\perp = 0.1, 0.05, 0.02, 0.01$ G, 
$f_u = 0.50$ to $0.95$ in $0.05$ steps, and 
$\gamma_l = 3,4,5,6,7,8,9,10$.  We realized that the shock strength,
$k$, was not a physically meaningful parameter in this case, so we 
set $k = 1.0$ for our fine grid. Out of these 900$+$ models, 
we found 4 models meeting the plausible criteria discussed 
above.  All of these models had $\ell = 1.0$, 
$B_\perp = 0.05$G or $0.1$G, $f_u = 0.7-0.9$, and 
$\gamma_l = 5$ or $6$.  The best overall model from this set
is plotted in figure \ref{f:spec_compD_mod1cd}.

The model plotted in Figure \ref{f:spec_compD_mod1cd} 
has $\ell = 1.0$, $B_\perp = 0.05$ G,
$f_u = 0.8$, and $\gamma_l = 6$.  While the comparison between
model and data was done at 15 GHz and above, we show here
the full spectra of the model at all our observing 
frequencies.  Note that the model deviates from the 
fractional linear polarization and EVPA data below 15 GHz. 
This trend could be easily explained if 
the amount of internal rotation reduced more quickly than
expected at larger $r$, perhaps due to local conditions
in the jet or an increase in the
lower cutoff to the relativistic particle spectrum,
$\gamma_l$, with $r$.  This deviation illustrates 
an important limitation of this model which can only
simulate broad spectral trends from smoothly varying
physical properties.  

For the circular polarization at 15 GHz and below in 
Figure \ref{f:spec_compD_mod1cd}, we
have added a dotted line which includes the contributions 
from the rotation-driven conversion models of 
components 5 and 4, given in figure \ref{f:spec_comp54_pol}.  
Note that the total circular polarization at low frequency 
is approximately
correct with this addition, although the total falls
a bit higher than the measured 8.0 GHz value.

The intrinsic magnetic field direction in this model 
corresponds to a position angle of $-129^\circ$, which 
places it between the structural position angles of 
components 5 and 4.  This is consistent 
with a parallel magnetic field ordered along a jet axis which
points at $\simeq -129^\circ$.  All of our plausible models gave
de-rotated magnetic field directions between $-129^\circ$ and
$-133^\circ$ to give the best match to the 15, 22, and 24 GHz data.
Given that all of these models are dominated by vector-ordered
field along the jet axis, i.e. $f_u \geq 0.7$, this is an
important consistency check, revealing that not only does 
this model produce the appropriate amounts of both linear and
circular polarization, but it also produces approximately
the right amount and sign of EVPA rotation for the 
emerging radiation. 

To check our dependence on the power-law exponents, $m$ and $n$,
we repeated the coarse grid described above for
two additional combinations, $m=1.3$, $n=3.4$ and $m=2.3$, $n=2.3$,
both of which match our Stokes $I$ spectrum.  We found only one
model which fit our plausibility criteria, and it had $m=2.3$, $n=2.3$,
$\ell = 1.0$, $B_\perp = 0.02$ G, $f_u = 0.8$, $\gamma_l = 10$, 
and $k = 0.7$.  The intrinsic magnetic field direction in this 
model corresponds to a position angle of $-128^\circ$.
These results present a very similar physical picture
to the best models with $m=1.8$ and $n=2.8$.  

\section{Discussion}
\label{s:discuss}

\subsection{Homogeneous Components 5 and 4}
\label{s:discuss_comp54}

For components 5 and 4, we find little to no circular polarization
at 22 and 24 GHz and as much as $0.5$\% to $1.0$\% circular 
polarization by 8.0 GHz.  In \S{\ref{s:specP_comp54}}, we explored two
possible physical models to explain this circular polarization
spectra both based on Faraday conversion of linear polarization
into circular.  The first model corresponds to model (3) described
in section \S{\ref{s:models}}, {\em rotation-driven conversion}, and
involves linear polarization generated by a shocked magnetic field, a
vector-ordered component of field along the jet axis which causes
Faraday rotation of the linear polarization within the jet, and
finally conversion of the rotated linear polarization into circular
polarization by the shocked magnetic field.  The results of this model
were plotted in Figure \ref{f:spec_comp54_pol}.  The second model
corresponds to model (5) described in section \S{\ref{s:models}},
conversion in a helical field, and involves linear polarization 
generated at the back of the jet being converted into circular
polarization by the magnetic field at the front of the jet which
is at some angle with respect to the field at the back of the
jet.

Both of the models for components 5 and 4 described above are
Faraday conversion models, as this will give the necessary steep
spectrum, and both explain the circular polarization of these
components adequately.  However, we prefer the rotation-driven
conversion model because it not only explains the circular
polarization but also the linear polarization, including
the amount and sign of the observed Faraday rotation.  As
described in \S{\ref{s:specP_comp54}}, in this model the
amount and sign of the observed circular polarization is 
essentially fixed by the amount of internal Faraday rotation
along with the observed degree of magnetic field order given
by the linear polarization.  On the other hand, the helical
field model has greater freedom to tune the amount and
sign of the observed circular polarization because it
decouples the observed Faraday rotation into an external
screen which may have different properties than the jet
itself.  

If all of the observed Faraday rotation in components 5
and 4 is internal to the jet, the Faraday conversion depth
at 8.0 GHz, $\tau_c$, is constrained to lie in the ranges 
$1.5-3.0$ and $0.75-1.5$ for components 5 and 4 respectively
to generate fractional circular polarization in the range
$m_c = 0.5 - 1.0$\% for each component.  
For the observed optical depths of these components, the
Faraday conversion depth 
constrains the Lorentz factor ratio between the emitting
particles and the low energy cutoff, $\gamma/\gamma_l$, to 
be in the range $20-400$ for component 5 and in the range $4.5-20$ for component 4.  
Using the estimated magnetic field strengths from \S{\ref{s:bfield_strength}} 
to find $\gamma$ at 8.0 GHz, the lower cutoff in the particle energy 
spectrum is constrained to be $1 \lesssim \gamma_l \lesssim 27$ and 
$9 \lesssim \gamma_l \lesssim 43$ for components 5 and 4 respectively.  Note 
that these limits scale with the estimated magnetic field strength 
like $1/\sqrt{B}$.  

From the observed Faraday rotation itself, we obtain only
a joint constraint on $f_u\ell \ln{\gamma_l}/\gamma_l^3$
as described in \S{\ref{s:specP_comp54}}.  If there is
significant cold ``thermal'' matter inside the jet, this 
constraint will be modified by an additional term: 
$+R_c(3/4)/\gamma_l$. 
where $R_c$ is the number ratio of cold to relativistic
particles in the jet \citep{JOD77}.

\subsection{Inhomogeneous Component D}

The inhomogeneous component D lies at the very base of the
parsec scale radio jet, and we argue in \S{\ref{s:specP_compD}}
that its Stokes $I$ and polarization spectra can be 
well modeled by a \citet{BK79} and \citet{K81} style conical 
jet with magnetic
field and particle strengths that scale as power-laws with
distance from the jet origin.  We prefer a physical model
for component D where the linear polarization is depolarized
and rotated by the same magnetic field and particles in the jet which
produce the Stokes $I$ emission, linear polarization, and
circular polarization.  These circumstances provide the
strongest constraints on the physical properties of the 
jet magnetic fields and particles as these properties alone
must produce consistent amounts and directions of both 
linear and circular polarization across multiple frequencies.
The physical model required by these constraints is
dominated by a vector-ordered magnetic field along the
jet axis with $f_u \geq 0.7$ which does three things: 
(1) produces the observed linear polarization, (2) 
generates internal Faraday rotation which rotates 
the EVPA and depolarizes the linear polarization to the
low fractional levels observed, and (3) 
produces the observed circular polarization largely 
as {\em intrinsic} circular polarization, although 
there is some contribution from {\em rotation-driven
conversion} of the linear polarization into 
circular.

It is interesting to note that the dominant
mechanism for producing circular polarization in
components 5 and 4 is rotation driven conversion,
not intrinsic circular polarization; however, these homogeneous 
components are not dominated by vector-ordered magnetic 
field along the jet axis which would be required 
for efficient intrinsic circular polarization 
production.  In fact, for components 5 and 4, some
fraction of vector-ordered field, $f_u$, is required 
to generate the necessary internal Faraday rotation, and 
indeed {\em the same polarity} of vector-ordered field is
required in components 5 and 4 as we deduce for component 
D.  

\subsubsection{Estimating the Jet Magnetic Flux}
\label{s:flux}

For our model of the inhomogeneous component D, we required 
the vector-ordered field to scale as $r^{-2}$ along the jet axis, 
consistent with conservation of magnetic flux in a conical jet.  
Continued conservation of magnetic flux further down the 
jet in components 5 and 4 would leave
some smaller fraction of vector-ordered field in these
components with the same polarity.  For components 5 and
4, it is not possible to compute the required magnetic flux 
because, as described in \S{\ref{s:discuss_comp54}}, 
$f_u$ is tied to other unknown parameters, and it is also 
not clear whether components 5 and 4, which are on different 
structural position angles, occupy the entire jet width at 
their location or represent bright spots within a jet with a 
wider opening angle.
However, these limitations don't apply to component D which
has $f_u \geq 0.7$, and we can estimate
the jet magnetic flux by fitting the transverse size of 
component D in our VLBA data.  We find the transverse size 
to be in the range $0.02-0.04$ mas; however, we note that we 
are at the very limits of our resolution here and can only use
this value to estimate the jet magnetic flux.  To convert from Gaussian 
FWHM to diameter of the jet, we multiply by $1.8$ \citep{M87}, and at a 
redshift $z = 0.536$, 1 mas corresponds to a linear scale of 
6.3 parsecs for standard cosmological parameters of 
$H_0 = 70$ km/s/Mpc, $\Omega_M = 0.3$ and $\Omega_\Lambda = 0.7$.  
For $f_u = 0.7$ and $B_\perp = 0.050$ G, the net magnetic flux in the
jet is in the range $\sim 2\times10^{34}-1\times10^{35}$ G cm$^2$.  

This net magnetic flux is, to our knowledge, the first 
estimate of the magnetic flux carried by a jet that takes proper 
account of the vector order of the magnetic field. In the absence 
of field line reconnection or entrainment, it is a conserved quantity 
and is therefore equal to the net poloidal magnetic flux at the central 
engine. If the field originates in the black hole magnetosphere, then 
the expected magnetic flux for an Eddington black hole is $\sim 2\times 10^{32}$ $M_8^{3/2}$ G cm$^2$ 
(Wardle \& Homan 2003, after correcting an arithmetic error; Begelman, Blandford \& Rees 1984). 
Setting this equal to the measured magnetic flux implies a black hole mass in the range 
$\sim 2\times 10^9 - 6\times10^9$ $M_\odot$ in 3C279. This would be at the top of the range of 
measured or inferred black hole masses found in the literature \citep{WU02} but is not
implausible; however, this estimate is an order of magnitude larger than the inferred 
black hole mass for 3C\,279 itself \citep{WU02, WLH04, LJG06}.  
Alternatively, the poloidal magnetic field may be 
anchored in the accretion disk, generally leading to larger magnetic fluxes 
because the area is much larger \citep[e.g.][]{BP82,VK04}. 
In the latter paper, the net magnetic fluxes in the two models presented are $10^{34}$  G cm$^2$ 
and $10^{35}$  G cm$^2$ (Vlahakis, private communication) which span our estimated range. 
The concordance between estimated and expected values suggests that circular polarization 
measurements can indeed measure a fundamental property of the central engine, and 
may even enable discrimination between models.

\subsubsection{Jet Composition}

The other physical parameter of interest  
constrained by this model is the lepton number $\ell$,
which we find to be $1.0$ in all of our plausible
models; however, the next smallest value we tested was
$\ell = 0.5$.  The lepton number is most strongly
constrained by the circular polarization which is largely due to 
instrinsic circular polarization in this model.  Smaller 
values of $\ell$ require larger values of $B$, such that 
$B \propto \ell^{-2}$, so if $\ell = 1.0$
for $B \simeq 0.05$ G as we find here, $\ell = 0.5$ would correspond
to $B \simeq 0.2$ G.\footnote{Note that the dependence of $B$ on $\ell$
is further complicated by the fraction of uniform field, $f_u$, such that
the quantity $f_u\ell B^{0.5}$ is approximately constant for $f_u$ near unity;
however, as $f_u$ is already $\geq 0.7$ not much additional
reduction in $\ell$ can be gained by increasing $f_u$.}  
Pushing this further, values as small as 
$\ell = 0.1$ might be allowed for stronger magnetic fields up to a  
few Gauss; however, larger values of $B$ produce correspondingly
larger values of the magnetic flux which is already quite large
when estimated assuming $B = 0.05$ G, see \S{\ref{s:flux}}.  Additionally,
from our observed flux density, angular size, and estimated optical depth 
at 22 GHz, we estimated $B\simeq 0.04$ G in \S{\ref{s:bfield_strength}}, and while differing from
this value by a factor of a few is certainly possible, the large deviations 
required for $\ell = 0.1$ or less are implausible. We therefore find 
that in this model, the jet in 3C279 is at least dynamically dominated by 
protons with $\ell$ likely to be greater than or equal to about $0.5$, 
corresponding to $\gtrsim 75$\% of the radiating particles being electrons.   

In the first parsec-scale circular polarization study of 3C\,279, 
\citet{WHOR98} preferred a jet dominated by 
electron-positron pairs, and their argument was based on 
their conclusion that $\gamma_l \leq 20$ and therefore
a jet dominated by electron-positron pairs was required
by kinetic luminosity arguments originated by \citet{CF93}.
Applying the same calculation as \citet{WHOR98} for jet energy flux 
to these observations, we derive a similar constraint with $F_E = 
1.33\times10^{45}[1+153\ell/\gamma_l]$ ergs/s.  This calculation
assumes a magnetic field strength of $B_\perp = 0.05$ G and,
like that of \citet{WHOR98}, assumes equipartition between the
magnetic field and particle energies.  Therefore, this calculation 
should be taken as a lower limit on the jet kinetic luminosity when
the field strength is $0.05$ G.  \citet{WHOR98} found a value for
$F_E$ about twice this, and they concluded that $\gamma_l \leq 20$ required
a mostly electron-positron jet to avoid carrying much more energy than appears to 
be dissipated on larger scales.  It should be noted that in
our model here, because the field strength, $B$, scales roughly with
$\ell^{-2}$, reducing $\ell$ far below 1.0 does not reduce the energy
carried by the jet, but rather increases it sharply as $\ell^{-4}$.

It seems that in this picture for component D there are only two 
ways to reduce the energy carried by the jet in the calculation given above.  
The first is reducing the magnetic field strength, which is possible as 
we had at least one model with $B_\perp = 0.02$ G when we tried power-law 
indices of $m=2.3$ and $n=2.3$, as described in \S{3.4.2}.  The second
is by increasing $\gamma_l$.  As we noted in the introduction, work by 
\citet{BF02} and \citet{RB02} showed that the addition of thermal plasma to the jet 
generates enough internal Faraday rotation to allow larger values
of $\gamma_l$ therefore relaxing the constraint on $\gamma_l$ found
by \citet{WHOR98}.  In the model for component D explored here, a
similar ambiguity exists for the value of $\gamma_l$.  All
of our plausible models have $\gamma_l \leq 10$; however, this
constraint on $\gamma_l$ is driven by the need to generate
enough internal Faraday rotation and depolarization by the
relativistic plasma itself and the addition of thermal matter 
to the jet would relax this constraint on $\gamma_l$.  Our
models for components 5 and 4 provide separate limits of 
$5 \lesssim \gamma_l \lesssim 35$, based on Faraday conversion, that
are not subject to this ambiguity.  Taken together, the uncertainties
in the value of $B_\perp$ and $\gamma_l$ allow for smaller values
of $F_E$, perhaps an order of magnitude smaller than that calculated
by \citet{WHOR98}.

\subsubsection{Uniqueness of Our Model}

The model described for component D in the previous sections
is of an inhomogeneous jet which is dominated by vector-ordered,
poloidal field along the jet axis. In this model, the poloidal 
magnetic field is responsible for generating the observed circular 
polarization as intrinsic circular polarization, and the same field 
provides the internal Faraday rotation and depolarization which produces the 
relatively low levels of observed linear polarization with a distinctly
rotated EVPA which is offset from the jet axis.  The inhomogeneous
nature of the component explains the Stokes $I$ spectrum, the circular
polarization spectrum, and the approximately flat spectrum for the
fractional linear polarization and its EVPA, despite the large Faraday
depth required.  It is these properties that we consider to be robust
aspects of our model, and they lead naturally to our constraints on
the composition of the radiating particles as parameterized by 
lepton number $\ell$.  

We expect that the details of the inhomogeneous model we have used 
are not unique, and that other inhomogeneous models could reproduce 
our observations just as well if they have the essential properties
listed above.  Indeed, as discussed in \S{3.4.2},
any inhomogeneous model with smoothly varying magnetic field and particle
densities is itself an idealization of the base of the jet which likely 
has local variations in these properties.  In this respect, we see our
detailed numerical results from \S{3.4.2} only as estimations of the
physical properties of this feature.

Finally, we reiterate that if an external Faraday screen is 
responsible for rotating and depolarizing the linear polarization, 
a wider range of general magnetic field and particle models could produce
the observed circular polarization. Parsec-scale Faraday rotation measure 
observations of 3C279 have been published by \citet{T98,T00} and 
\citet{ZT01, ZT03, ZT04} 
spanning five epochs from early 1997 through mid-2001. Their results
agree with ours in that they find consistently negative, but variable, 
Faraday rotations in the core region of magnitudes encompassing those
we see in components D, 5, and 4.  The variable rotation measures
they observe fit naturally into our model where the amount of
poloidal magnetic field drops off sharply from the core region, 
and newly emerging components can sample a range of internal 
rotation measures as they propagate down the jet.  However, an
external screen, very close to the jet, could also have these 
properties. 
Further out in the jet, these authors find a strong jet component to 
have a small amount of negative Faraday rotation at about 3 milli-arcseconds from
the core in their 1997 and 1998 epochs; however, by 2000, this
component shows a small amount of positive Faraday rotation, 
less than $+50$ rad/m$^2$ in the observer's frame of reference.  
This shift to positive rotation measure could be related to a collision
and re-alignment event this component underwent in mid-1998
\citep{H03}, or the component could simply be far enough from
the central engine that other external Faraday screens between us 
and the source are playing the dominant role.\footnote{Note that \citet{ZT04} 
find an integrated rotation measure of $+31$ rad/m$^2$ for 3C279 which is
nearly an order of magnitude smaller than even the smallest negative 
rotation measure we find here for component 4, and thus correction
for this integrated value would have essentially no impact on our 
results.}   

This evidence indicates that the negative Faraday rotation observed
near the core region by ourselves and others is indeed produced local
to the jet. It could be due to the fields internal to the jet, as we 
propose here.  Alternatively, it could be generated externally, but 
close to the jet, by fields and particle densities which scale with 
frequency and position in just the right way to produce both the 
approximately constant fractional linear polarization {\em and} 
offset EVPA observed in component D at all frequencies.  
We regard the fact that linear polarization of this kind comes out 
naturally from the internal rotation model as strong evidence in 
favor of it.

\subsection{Comparison to Integrated UMRAO Monitoring}

An interesting result from the integrated UMRAO monitoring
of 3C\,279 at 4.8, 8.0, and 14.5 GHz is that 3C\,279 shows
occasional sign reversals in its circular polarization 
at 4.8 GHz but not at higher frequencies \citep[][and Figure 5]{AAH06}.  
An important
question is whether these sign reversals can be understood
in the context of the model of the core region preferred
for the VLBA data.  From Figure \ref{f:specI} it is clear
that the core region flux at 4.8 GHz is dominated by
components 5 and 4.  The Faraday conversion models preferred
for components 5 and 4 will naturally produce negative
circular polarization at high optical depths, at a frequency
about one-third of the spectral turnover \citep{JOD77}.
With spectral turnovers near 9.0 GHz, components 5 and 4
should not be producing negative circular polarization
at 4.8 GHz in the VLBA epoch investigated here, and indeed
negative circular polarization is not observed near
this epoch; however, when these components were first
originating near the base of the jet, they must have been
more compact with higher turnover frequencies and may
well have generated the negative circular polarization
observed at 4.8 GHz in mid-2005 along with the increase
in Stokes $I$ at that time.  Strong negative circular 
polarization is also seen at 4.8 GHz at several other epochs
during this period: early 2004, mid-2004, mid-2006, 
late 2006, and late 2007.  Each of these events also appear 
to be linked with increases in Stokes $I$, suggesting that 
new components may be ejected from the base of the jet at 
these times.

\section{Conclusions}
\label{s:conclude}  

We have made multi-frequency, parsec scale observations
of the core region of 3C\,279 in Stokes $I$, linear
polarization, and circular polarization.  The core region 
of 3C\,279 consists of three main components: two 
homogeneous components, 5 and 4, and the inhomogeneous 
component D at the base of the jet, and we have modeled 
the full polarization spectra of these components with
radiative transfer simulations to constrain the
magnetic field and particle properties of the 
parsec-scale jet in 3C\,279.

We find that the polarization properties of the core
region, including the amount of linear polarization, the
amount and sign of Faraday rotation, and
the amount and sign of circular polarization
can be explained by a consistent physical model.  The
base of the jet, component D, is modeled as a conical
inhomogeneous Blandford-K\"onigl style jet \citep{BK79,K81}. 
The magnetic field of this feature is dominated by vector-ordered
poloidal magnetic field along the jet axis, and we estimated 
the net magnetic flux of this field.  This poloidal field at the
base of the jet produces intrinsic circular polarization and depolarizes 
and rotates the linear polarization.  In the homogeneous components 5
and 4, further down the jet, the magnetic field is
dominated by local shocks, and a much smaller fraction
of vector-ordered, poloidal field remains along the jet axis.  
This remaining vector-ordered field provides a consistent 
internal Faraday rotation, which allows Faraday conversion
to generate the appropriate amount and spectra of circular 
polarization from these components.

We note that this physical model is not unique if one
allows the observed Faraday rotation and depolarization
to occur in screens external to the jet.  Such external
screens would de-couple the linear and circular polarization
and allow a wider range of physical parameters; however, we find
the fact that all of the essential polarization characteristics can be
simultaneously produced by magnetic field and plasma 
properties internal to the jet itself to be compelling 
motivation for this model.  

With this model, we can additionally constrain
the particle properties of the jet.  We find the jet
to be kinetically dominated by protons with a lepton
number $\ell \gtrsim 0.5$ corresponding to $\gtrsim$ 75\%
of the radiating particles being electrons, and therefore we 
cannot rule out a significant admixture of positrons.
Based on the amounts of Faraday conversion deduced
for the homogeneous components 5 and 4, we find
a plausible range for the lower cutoff in the relativistic
particle energy spectrum to be 
$5 \lesssim \gamma_l \lesssim 35$.

\acknowledgments

This research was supported by an award from Research Corporation, 
NSF grants AST-0707693, AST-0607523 and AST-0406923, and by 
funds from the University of Michigan.  We thank the referee
for insightful comments that have helped us clarify our methods
and results.


\appendix

\section{Opacity Effects on Linear Polarization}

Here we explore the effect of optical depth, $\tau$, on the observed 
fraction of linear polarization, $m_l$, from a 
partially ordered magnetic field.  Our goal is to
obtain an analytic model for $m_l$ as a function of 
$\tau$ which can be used in fitting spectra of 
homogeneous synchrotron sources. We take the
ordered part of the magnetic field to be in the
plane of the sky at a position angle of $90^\circ$ 
so that the emitted linear polarization is entirely
Stokes-$Q$.  We also assume no internal Faraday rotation
or conversion.  Under these circumstances, the
equations of radiative transfer reduce to the following
expressions \citep[e.g.][]{Jones88}.

For a completely ordered magnetic field, we have the
exact differential equations ...

\begin{equation}
\frac{dI}{d\tau} + I + \zeta_Q Q = J 
\end{equation}
\[\frac{dQ}{d\tau} + Q + \zeta_Q I = \epsilon_Q J\]

\noindent where $\epsilon_Q = 0.6923$ and $\zeta_Q = 0.7500$ are the 
emission and absorption coefficients respectively for 
$\alpha = -0.5$.  For optically thin emission in a completely
ordered magnetic field, the fractional 
linear polarization is equal to the emission coefficient: 
$m_{l} (\tau = 0) = \epsilon_Q$.  

For a partially ordered field, we will substitute appropriately
scaled emission and absorption coefficients, $\epsilon$ and $\zeta$,
for $\epsilon_Q$ and $\zeta_Q$ respectively.  

\begin{equation}
\frac{dI}{d\tau} + I + \zeta Q = J
\end{equation}
\[\frac{dQ}{d\tau} + Q + \zeta I = \epsilon J\]

The scaled emission
coefficient $\epsilon$ must give the observed fractional linear
polarization at $\tau = 0$, so $\epsilon = m_{l} (\tau = 0)$, and 
we can guess that the appropriate scaling for the absorption 
coefficient is similar.  We parameterize 
$\zeta = \zeta_Q \left( \frac{m_l (\tau = 0)}{\epsilon_Q} \right)^\eta$,
where $\eta$ is an unknown power, likely to be close to unity.
  
These coupled, first order differential equations can be
easily solved and we find the resulting fractional linear
polarization as a function of opacity.

\begin{equation}
\label{e:lp_model}
m_l = \frac{Q}{I} = \frac{(1+\epsilon)(1-\zeta)(1-e^{-(1+\zeta)\tau})-
                          (1-\epsilon)(1+\zeta)(1-e^{-(1-\zeta)\tau})}
                         {(1+\epsilon)(1-\zeta)(1-e^{-(1+\zeta)\tau})+
                          (1-\epsilon)(1+\zeta)(1-e^{-(1-\zeta)\tau})}
\end{equation}

In the limit $\tau \rightarrow 0$, $m_l = \epsilon$ as we expect, and as 
$\tau \rightarrow \inf$, $m_l = \frac{\epsilon-\zeta}{1-\zeta\epsilon}$. 
All of these expressions are exact for a completely ordered magnetic field;
however, we are interested in a partially ordered field using the scaled
versions of the emission and absorption coefficients given above: 
$\epsilon = m_l(\tau=0)$
and $\zeta = \zeta_Q \left( \frac{m_l (\tau = 0)}{\epsilon_Q} \right)^\eta$.

\begin{figure}
\figurenum{A1}
\begin{center}
\includegraphics[scale=0.45,angle=0]{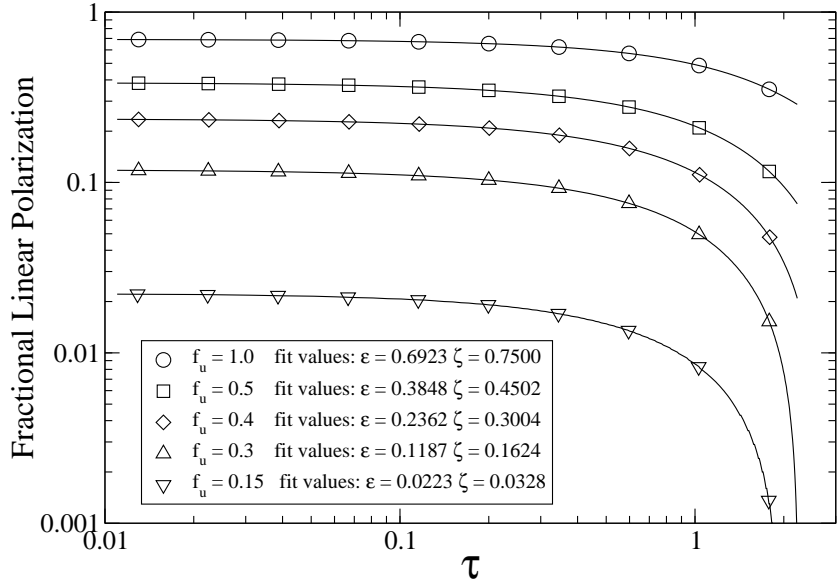}
\end{center}
\vspace{0.2in}
\figcaption[]{\label{f:lp_test}
Fractional linear polarization plotted against optical depth from 
our radiative transfer simulation for a partially ordered magnetic
field.  The magnetic field is taken to be in the plane of the sky
with fractional order given by the parameter $f_u$ described in 
\S{\ref{s:rad_model}}.  The results of the simulation are plotted
as open symbols.  A fit to the analytical expression given by 
equation \ref{e:lp_model} is plotted as a solid line for each 
simulation, and the fitted values for $\epsilon$ and $\zeta$
appear in the figure legend.
}
\end{figure}

To test these expressions, we ran the radiative transfer simulation described
in \S{\ref{s:rad_model}} with $\alpha = -0.5$ for various fractions, $f_u$, for the 
vector-ordered magnetic field, taken to be in the plane of the sky.  We used 
$N=10^6$ cells along the line of sight
to minimize any net contribution from the random component of magnetic field
in each cell.  We also set the lepton number $\ell = 10^{-5}$ to eliminate the
effects of internal Faraday rotation.  The fractional polarization, $m_l$ from these 
runs is plotted against optical depth, $\tau$, in figure \ref{f:lp_test}.
The model given in equation \ref{e:lp_model} is plotted for each value of 
$f_u$ where the best values of $\epsilon$ and $\zeta$ were found by a least-squares
fit.  The best-fit values for $\epsilon$ and $\zeta$ are given in the
legend of figure \ref{f:lp_test}, and we can see that indeed $\epsilon = m_l(\tau=0)$
as expected, and we find that our scaled value for 
$\zeta = \zeta_Q \left( \frac{m_l (\tau = 0)}{\epsilon_Q} \right)^\eta$ is an 
excellent approximation if $\eta = 0.86$. 

An additional factor to consider is the possibility of depolarization due
to internal Faraday rotation.  \citet{B66} predicts additional depolarization
of $\sin(\Phi)/\Phi$ where $\Phi_{Burn} = 2.0\Delta\chi$; however, this is for 
a purely optically thin case.  Figure A2 shows the results of our radiative
transfer simulation with $f_u = 0.3$ with the jet oriented at $\theta = 45^\circ$ 
to the line of sight to generate internal Faraday rotation.  The amount of
Faraday rotation as a function of optical depth was chosen to be similar to
what we observe for components 5 and 4, with $\lesssim 40^\circ$ of internal
rotation up to an optical depth of $\tau\sim 1$.  Note that the internal
rotation is well fit by a $\lambda^2$ regression up to $\Delta\chi \simeq 40^\circ$.  
Panel (b) of this figure shows the additional effect of internal depolarization
beyond the pure optical depth reduction to $m_l$.  To include the effect of 
internal Faraday depolarization in these plots, we used the linear regression fit  
to $\Delta\chi = RM \lambda^2$ to produce the $\Phi$ values.  Note that $\Phi_{Burn}$ 
somewhat over-predicts the effect of internal depolarization, and we found that
a modified factor $\Phi = 1.7\Delta\chi$ produced a better match to the simulation
data.  

Taken together, equation A3 and above depolarization rule for internal 
Faraday rotation, allow us to include fractional linear polarization, $m_l$,
in our analytical fit to the observed Stokes $I$ spectrum of homogeneous
jet components to find their peak frequency.  To test this technique, we
ran this procedure on a detailed numerical simulation of component 5 from
our radiative transfer program.  We took the results of this simulation
at each frequency and treated them as data with the same uncertainties 
on $m_l$ and Stokes $I$ as the real data from component 5.  The simulation
had $f_u = 0.05$ with the jet oriented at $\theta = 1.5^\circ$ to the line
of sight with a bulk Lorentz factor of $\Gamma=15$.  To generate the right
amount of fractional linear polarization at high frequency a shock of
strength $k = 0.64$ was applied and the lower cutoff to the electron 
energy spectrum was set to $\gamma_l = 19$ to generate the observed
amount of Faraday rotation at each frequency.  The observed peak frequency
of the component was set to be $\nu_{peak} = 8.95$ GHz with a spectral index
of $\alpha = -0.50$.  We then ran Stokes $I$ and $m_l$ values from this
simulation through our analytical peak fitting program and found 
$\nu_{peak} = 8.80$ GHz with $\alpha = -0.49$, only $2$\% less than the input 
values of the simulation.

\begin{figure}
\figurenum{A2}
\begin{center}
\includegraphics[scale=0.75,angle=0]{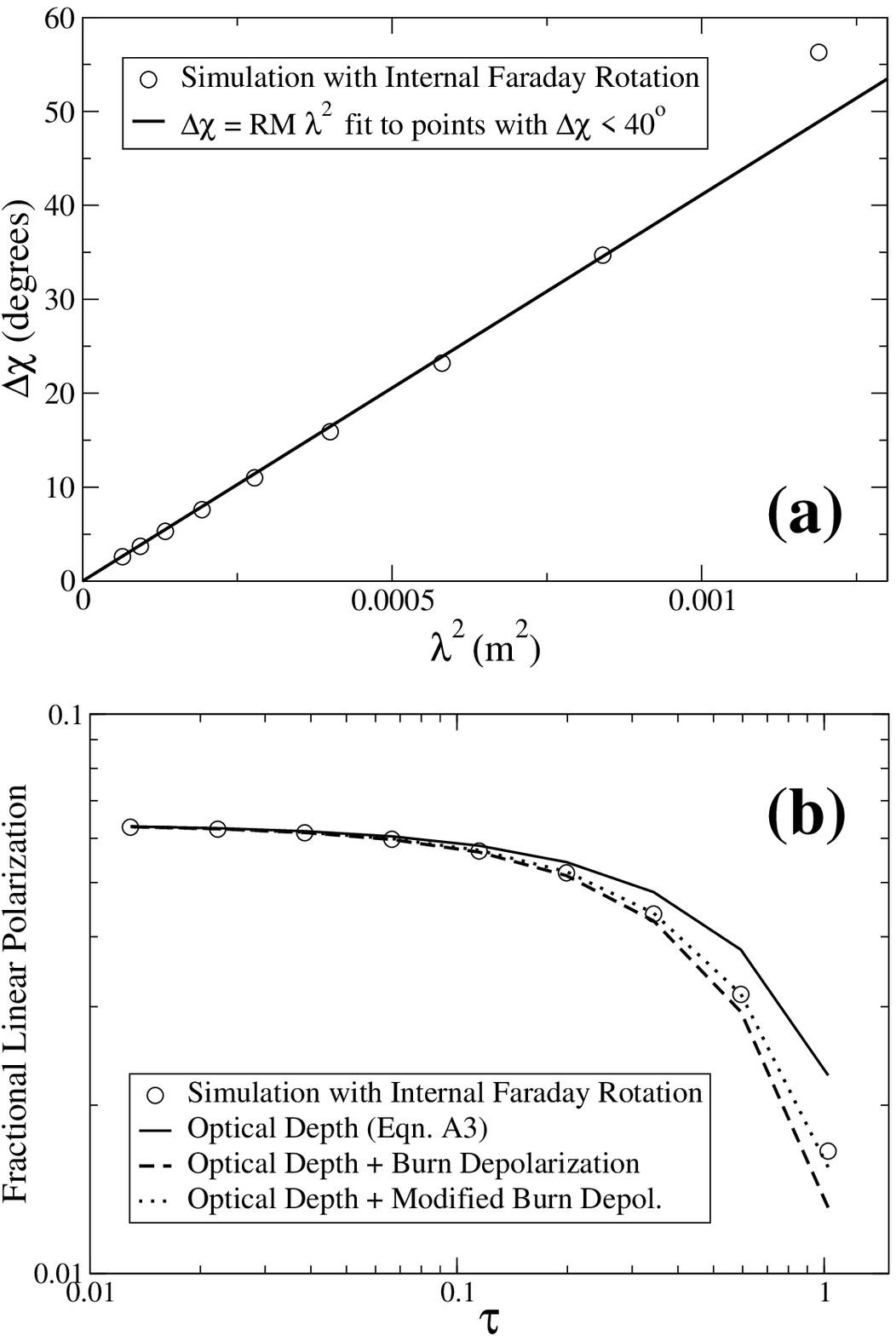}
\end{center}
\figcaption[]{\label{f:lp_test2}
Panel (a) is observed polarization position angle from our 
radiative transfer simulation with a partially ordered field and
internal Faraday rotation.  The overplotted line is a linear
regression to all of the points with total rotation less than
$40^\circ$. Panel (b) plots fractional linear polarization against 
optical depth from the same simulation for all points with total 
rotation less that $40^\circ$.  The overplotted lines are the 
effects of pure optical depth (solid line), optical depth plus
Burn style depolarization (dashed line), and optical depth plus
the modified Burn depolarization described in the text (dotted line).
}
\end{figure}



\end{document}